\documentclass[a4paper]{article}

\usepackage[english]{babel}
\usepackage[T1]{fontenc}
\usepackage[affil-it]{authblk}
\usepackage{etoolbox}
\usepackage{lmodern}

\makeatletter
\patchcmd{\@maketitle}{\LARGE \@title}{\fontsize{14}{19.2}\selectfont\@title}{}{}
\makeatother

\usepackage[a4paper,top=3cm,bottom=2cm,left=3cm,right=3cm,marginparwidth=1.75cm]{geometry}
\usepackage[english]{babel}
\usepackage[autostyle]{csquotes}

\usepackage{amsmath}
\usepackage{graphicx}
\usepackage[colorinlistoftodos]{todonotes}
\usepackage[colorlinks=true, allcolors=blue]{hyperref}

\date{}
\newcommand{\ignore}[1]{{}}

\usepackage[utf8]{inputenc}

\usepackage{pgfplots} 
\usepackage{tikz}
\usetikzlibrary{shapes,backgrounds,shapes,positioning,petri,topaths}

\usepackage{graphicx}
\usepackage{tikz}
\usepackage{tkz-graph}
\usepackage[labelfont=bf, font=small, labelsep=period, skip=0pt]{caption}

\usepackage{multirow}

\usepackage[font=scriptsize,labelfont=scriptsize]{subfig}
\usepackage{tikz}
\usetikzlibrary{backgrounds,fit,decorations.pathreplacing, shapes}  
\usepackage{tkz-graph}

\tikzstyle{vertex}=[circle,draw=black, fill=white,sloped,minimum size=17pt,inner sep=5pt]
\newcommand{\vertex}{\node[vertex]}

\usepackage{tikz}
\usetikzlibrary{decorations.pathreplacing,calc}

\usepackage{algorithm} 
\usepackage{algpseudocode}
\usepackage{adjustbox}
\usepackage[utf8]{inputenc}

\usepackage{listings}

\lstdefinestyle{mystyle}{
    basicstyle=\footnotesize,
    breakatwhitespace=false,         
    breaklines=true,                 
    captionpos=b,                    
    keepspaces=true,                 
    numbers=left,                    
    numbersep=5pt,                  
    showspaces=false,                
    showstringspaces=false,
    showtabs=false,                  
    tabsize=2
}
\newcommand{\figref}[1]%
{Fig. \ref{#1}%
}

\newcommand{\tableref}[1]%
{Table \ref{#1}%
}

\newcommand{\algorithmref}[1]%
{Algorithm \ref{#1}%
}

\newcommand{\sectionref}[1]%
{Section \ref{#1}%
}

\newcommand{\lineref}[1]%
{Line \ref{#1}%
}

\pgfplotsset{compat = newest}

\def\code#1{\texttt{#1}}

\usepackage[sort&compress,numbers]{natbib}

\usepackage{authblk}

\title{\textbf{Dynamic Laplace: Efficient Centrality Measure for Weighted or Unweighted Evolving Networks}}

\author[1, 2]{Mário Cordeiro}
\author[1, 2]{Rui Portocarrero Sarmento}
\author[2]{Pavel Brazdil}
\author[2]{João Gama}

\affil[1]{Doctoral Program in Informatics Engineering, Faculty of Engineering, University of Porto}
\affil[2]{INESC TEC - Institute for Systems and Computer Engineering, Technology and Science}

\begin{document}
\maketitle




\maketitle

\begin{abstract}
With its origin in sociology, Social Network Analysis (SNA), quickly emerged and spread to other areas of research, including anthropology, biology, information science, organizational studies, political science, and computer science. Being it's objective the investigation of social structures through the use of networks and graph theory, Social Network Analysis is, nowadays, an important research area in several domains. Social Network Analysis cope with different problems namely network metrics, models, visualization and information spreading, each one with several approaches, methods and algorithms. One of the critical areas of Social Network Analysis involves the calculation of different centrality measures (i.e.: the most important vertices within a graph). Today, the challenge is how to do this fast and efficiently, as many increasingly larger datasets are available. Recently, the need to apply such centrality algorithms to non static networks (i.e.: networks that evolve over time) is also a new challenge. Incremental and dynamic versions of centrality measures are starting to emerge (betweenness, closeness, etc). Our contribution is the proposal of two incremental versions of the Laplacian Centrality measure, that can be applied not only to large graphs but also to, weighted or unweighted, dynamically changing networks. The experimental evaluation was performed with several tests in different types of evolving networks, incremental or fully dynamic. Results have shown that our incremental versions of the algorithm can calculate node centralities in large networks, faster and efficiently than the corresponding batch version in both incremental and full dynamic network setups.
\end{abstract}

\section{Introduction}

Centrality measures in Social Network Analysis (SNA) have been an important area of research as they help us to identify the relevant nodes. Researchers have invested a lot of effort to develop algorithms that could efficiently calculate the centrality measures of nodes in networks. With the explosion of social networks users, for example, like Twitter or Facebook, the networks have grown to the point where the use of batch algorithms cannot handle this data efficiently. Thus, to perform the analysis of large and changing networks, it is necessary to adopt streaming techniques and use of incremental algorithms. This way, researchers try to speed-up the process and use less memory whenever possible, by avoiding to process the full network in each iteration.

Our contribution, in this paper, is an efficient solution to calculate a particular centrality measure, the Laplacian centrality, in an \textit{incremental only} or \textit{full dynamic} setting. By \textit{incremental only} we mean networks in which only new nodes and edges are added to the network in subsequent time snapshots, by \textit{fully dynamic} we mean support for addition and removal of nodes and edges in subsequent time snapshots. We present a solution that is accurate and faster than the corresponding batch algorithm for Laplacian centrality on large evolving networks. The proposed incremental algorithm supports both weighted or unweighted networks.

Succinctly, this document starts with an introduction to related work in \sectionref{rel}. After the introduction to the related work, we briefly explain the nomenclature used throughout the paper in \sectionref{Num}. We explain our incremental algorithm in \sectionref{IncrementalLaplaceCentralityalgorithm}. Then, in \sectionref{Res}, we write about the results of our experiments with the developed algorithm. Finally, in \sectionref{Conc}, we conclude our work and write about possible directions for future action regarding the area covered in this document.

\section{Nomenclature}\label{Num}

The undirected unweighted graph that represents a network with $N$ nodes and $M$ links is given by $G = (V,E)$. The components of the graph are: the node set $V$, which is just a list of indices; the edge set $E$ where each edge consists of two vertices. The undirected graph has $n = \vert V\vert$ nodes, $V=\{u_1,u_2, ..,u_n\}$, and $e = \vert E\vert$ edges, $E=\{(i_1,j_1),(i_2,j_2), ..,(i_e,j_e)\}$.
For each node $u$, $d_u$ is its respective degree. A more strict type of graph is the so-called directed graph (or directed network). Directed graphs, can be defined as graphs whose all edges have an orientation assigned, so the order of the vertices they link matters. Formally, a directed graph $D$ is an ordered pair ($V(D)$, $A(D)$) consisting of a nonempty set $V(D)$ of vertices and a set $A(D)$, disjoint from $V(D)$, of arcs. If $e_{ij}$ is an arc and $u_i$ and $u_j$ are vertices such that $e_{ij} = (u_i, u_j)$, then $e_{ij}$ is said to join $u_i$ to $u_j$, being the first vertex $u_1$ called initial vertex, and the second vertex $u_j$ called the terminal vertex. For undirected and unweighted graphs, adjacency matrices are binary (as a consequence of being unweighted) and symmetric (as a consequence of being undirected, meaning that $a_{ij} = a_{ji}$), with $a_{ij} = 1$ representing the presence of an edge between vertices $u_i$ and $u_j$, and $a_{ij} = 0$ representing the absence of an edge between vertex pair ($u_i$,$u_j$). For undirected and weighted graphs, the entries of such matrices take values from the interval [0, $max(w)$] and are symmetric. For directed and weighted graphs, the entries of such matrices take values from the interval [0, $max(w)$] and are non-symmetric. In any of these cases, we deal with non-negative matrices.

\section{Related Work}\label{rel}

\subsection{Centrality Measures for Static Networks}

\begin{figure}[H]
    \centering
    \includegraphics[scale=0.25]{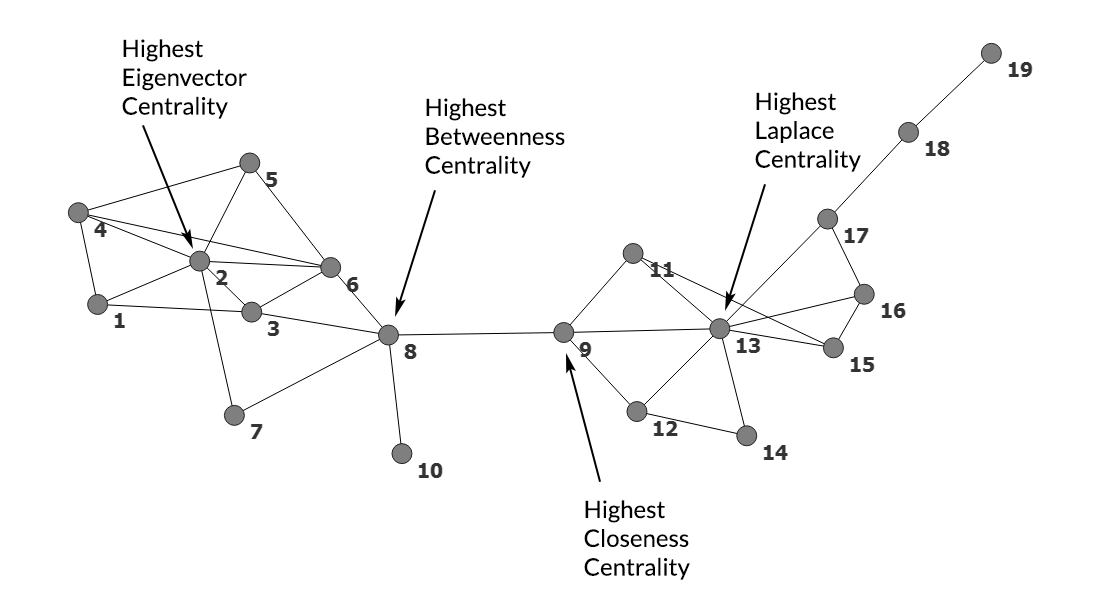}
    \caption{Different centrality measures in a graph}
    \label{fig:DifferentCentralities}
\end{figure}

In this section, we briefly introduce some of the commonly used centrality measures, in the context of retrieving centrality values for the nodes in a network. Although there is no consensual best centrality measure for graphs, several measures are accepted and give good centrality values that are valuable in different scenarios. \figref{fig:DifferentCentralities} show a comparison on the central node calculated via different methods. All these centrality measures are presented and explained below.

\begin{description}
    \item [\textbf{Betweenness Centrality:}] measures the extent to which a node lies between other nodes in the network. Thus, the nodes with higher betweenness are included in more shortest paths between nodes that are not directly connected. 
    Nodes with high betweenness occupy critical roles
    in the network structure since they usually have a
    network position that allows them to work as an interface
    between different groups of nodes \cite{DBLP:journals/widm/OliveiraG12}. The flow of information between two nodes that are not directly connected in the network, rely on the central nodes to propagate the information to these non-connected nodes \cite{wasserman1994social}. Thus, the central nodes, i.e., the nodes with higher values of this centrality measure are important nodes that provide better dissemination of information in the network. The Betweenness Centrality of node $u$ is given by:
    
    \begin{equation}
    \centering
    b_u=\sum _{s,t\epsilon V(G)\backslash u}{\frac {\sigma _{s,t} (u)}{\sigma _{s,t}}}
    \end{equation}
    
    where $\sigma _{s,t}$ denotes the number of shortest paths between
    vertices $s$ and $t$ (usually $\sigma _{s,t}$ = 1) and $\sigma _{s,t} (u)$ expresses the number of shortest paths passing through node $u$.\\
    
    \item [\textbf{Closeness Centrality}:] 
    is a measure based on the concept of distance to other nodes. A node has higher closeness centrality when the shortest path to the other nodes are shorter and lower closeness centrality when the shortest paths to all other nodes in the network are longer. Thus, Closeness Centrality is a measure of how fast a given node can reach all other nodes in the network, in average \cite{DBLP:journals/widm/OliveiraG12}. The Closeness Centrality of node $u$ is given by:
    
    \begin{equation}
    \centering
    Cl_u={\frac {N - 1}{\sum _{v\epsilon V(G)\backslash u} d(u,v)}}
    \end{equation}
    
    where $d(u,v)$ denotes the shortest paths between
    vertices $u$ and $v$ and $n$ expresses the number of nodes in the graph $G$. $N-1$ represents the the normalized form where $N$ is the number of nodes in the graph. For large graphs this difference becomes inconsequential so the $-1$ is dropped.\\
    
    \item [\textbf{Eigenvector Centrality}:] is given by the first eigenvector of the adjacency matrix. Therefore, eigenvector centrality assumes that the status of a node is recursively defined by the status of his/her first-degree connections, i.e., the nodes that are directly connected to a particular node \cite{DBLP:journals/widm/OliveiraG12}. Eigenvector Centrality of node $i$ is given by:
    
    \begin{equation}
    \centering
    x_i = {\frac {1}{\lambda }} \sum _{j = 1}^{n} a_{ij}x_j
    \end{equation}
    
    where $x_i/x_j$ denotes the centrality of node $i/j$, $a_{ij}$ represents
    an entry of the adjacency matrix $A$ ($a_{ij} = 1$ if
    nodes $i$ and $j$ are connected by an edge and $a_{ij} = 0$
    otherwise) and $\lambda$ denotes the largest eigenvalue of $A$.
    
    Eigenvector centrality is an important measure since it is based not only on the ammount of connections but also the quality of the links a node has.\\
    
    \item [\textbf{Laplacian Centrality}:] In \cite{Qi:2012:LCN:2181343.2181780}, Xingqin Qi et al. introduced a novel centrality measure. The authors stress that their new measure based on the Laplacian energy of a node would outperform Betweenness and Closeness centrality regarding complexity. Therefore, the authors achieved a more efficient way of retrieving measurements of centrality in a network. In their paper, authors also compare their new measure with Betweenness centrality and Closeness centrality and prove the reliability of their measure by providing similar results per node. The definition of the Laplacian Centrality of a node $u_i$ is based in the Laplacian Energy of the graph $G$ obtained after removing the node $u_i$ from the graph, and by following the expression \cite{Qi:2012:LCN:2181343.2181780}:
    
    \begin{equation}\label{eq3}
    \centering
    C_L (u_i, G) = \frac{(\Delta E)_i}{E_L (G)}=\frac {E_L (G) - E_L (G_i)}{E_L (G)}
    \end{equation}
    
    From equation \ref{eq3}, we can expect a Laplacian Centrality value between 0 and 1. These values increase correspondingly to the increased centrality of the node. Nonetheless, since the definition is in itself a normalization, it is a standard procedure to present the non-normalized values $(\Delta E)_i$ for each node. This way, we can achieve a faster-ranking computation without considering the total graph energy $E_L (G)$. We present adapted versions of the Laplacian Centrality in the scope of this work. Therefore, a detailed analysis of the algorithms is shown here. This is true either for the unweighted or the weighted variants, for further study and discussion within this document. In \algorithmref{BatchLaplaceCentralityAlgorithm}, method \code{LapCentUnweighted()}, is presented the implementation of the Laplacian Centrality measure as described in \cite{Qi:2012:LCN:2181343.2181780,Qi2013} for unweighted networks. Presented pseudocode is based on the implementation of the Laplacian Centrality as made available by \cite{Wheeler2015, igraph2014}. The weighted variant of the Laplacian Centrality is also made available by \cite{Wheeler2015, igraph2014} (see \code{LapCentWeighted()}). The main difference to the unweighted variant is the way how the centrality value is calculated for each node. For the weighted version, instead of considering only the degrees of the nodes, it needs to perform the calculation of all the two walks from affected nodes. This approach revealed to be unfeasible. Wheeler \cite{Wheeler2015} developed a different algorithm, the \algorithmref{BatchLaplaceCentralityAlgorithm}, \code{LapCentWeighted()}, that proven to be quicker, easy to implement and delivers good centrality results. For the calculation of each node Laplace Centrality, both unweighted and weighted methods are shown in \code{LapCentUnweighted()} and \code{LapCentWeighted()} methods, respectively. Both algorithms were adapted to work in an evolving network setting by processing each one of the temporal snapshots \code{Main()}, either incremental or fully dynamic. By observation of the \code{LapCentUnweighted()} method, for each loop of the algorithm, it can be concluded that the centrality parameter is a function of the local degree plus the degree of the neighbors (with different weights for each). Therefore, the metric is not a global measure. The local degree and the 1st order neighbors degree are all that is needed to calculate the metric for unweighted networks. The \code{Main()} method shows the pseudo-code required for performing batch Laplacian Centrality calculation on an evolving network. The algorithm will require performing a full calculation for each one of the snapshots $\{G_0, G_1, ...,G_n\}$ included in the dataset $\mathcal{D}_{ataset}$. Notice that no Laplacian Centrality data nor network data is shared between snapshots. We will explore this inefficiency of the algorithm in \sectionref{IncrementalLaplaceCentralityalgorithm} when proposing the incremental version of the algorithm.

    \begin{algorithm}[!th]
    \small
  \footnotesize
  \caption{Batch Laplace Centrality Algorithm}\label{BatchLaplaceCentralityAlgorithm}
  \begin{algorithmic}[1]
    \State $V\gets \{u_1,u_2, ..,u_v\}$ ,  $E\gets \{(i_1,j_1),(i_2,j_2), ..,(i_e,j_e)\}$

    \Procedure{LapCentUnweighted}{$G\gets (V,E)$} \Comment{for unweighted networks}
    
    \State$\mathcal{C}_{entralities} \gets \{\}$
    \State$\mathcal{D}_{egrees} \gets G.degrees()$
    
    \State$\mathcal{V}_{s} \gets G.nodes()$
    

    \For{\textbf{each} $v$ \textbf{in} $\mathcal{V}_{s}$}
    
        \State$\mathcal{N}_{eighbors} \gets G.neighbors()$
        \State$loc \gets \mathcal{D}_{egrees}[v]$
        \State$nei \gets 2 . \sum\limits_{i=1}^{\mathcal{N}_{eighbors}} \mathcal{D}_{egrees}[i]$
        \State$\mathcal{C}_{entralities} [v] \gets ({loc}^2 + loc + nei)$
      
    \EndFor
    
    \State\Return $\mathcal{C}_{entralities}$, $\mid\mathcal{V}_{s}\mid$

  \EndProcedure
  
  \Procedure{CW}{$G\gets (V,E)$, $\mathcal{V}_{ertex}$, $\mathcal{D}_{egrees}$} \Comment{centrality weight for weighted networks}
    
        \State$\mathcal{N}_{eighbors} \gets G.neighbors(\mathcal{V}_{ertex})$
        
        \State$cw \gets \sum\limits_{i=1}^{\mathcal{N}_{eighbors}} {(\mathcal{N}_{eighbors}[i].weight())}^2$
        
        \State$sub \gets \sum\limits_{i=1}^{\mathcal{N}_{eighbors}} [{(\mathcal{D}_{egrees}[i] - \mathcal{N}_{eighbors}[i].weight())}^2 - {(\mathcal{D}_{egrees}[i])^2}]$
        
        \State\Return $cw$, $sub$

    \EndProcedure

    \Procedure{LapCentWeighted}{$G\gets (V,E)$} \Comment{for weighted networks}
    
    \State$\mathcal{C}_{entralities} \gets \{\}$
    \State$\mathcal{D}_{egrees} \gets G.degrees()$
    
    \State$\mathcal{V}_{s} \gets G.nodes()$
    

    \For{\textbf{each} $v$ \textbf{in} $\mathcal{V}_{s}$}
    
        \State$loc \gets \mathcal{D}_{egrees}[v]$
        \State$cw$, $sub$ $\gets \Call{CW}{G, v, \mathcal{D}_{egrees}}$
        
        \State$\mathcal{C}_{entralities} [v] \gets ({loc}^2 - sub + 2*cw)$
      
    \EndFor
    
    \State\Return $\mathcal{C}_{entralities}$, $\mid\mathcal{V}_{s}\mid$

  \EndProcedure
  
  \Procedure{Main}{}
  
  \State $\mathcal{D}_{ataset} \gets \{G_{0}, G_{1}, ... ,G_{n}\}$
  
  \For{\textbf{each} $snapshot$ \textbf{in} $\mathcal{D}_{ataset}$} \Comment{calculation in the full network}
        \If{unweighted $\mathcal{D}_{ataset}$}
            \State$\mathcal{C}_{entralities}$, $\mathcal{N}_{um}\mathcal{C}_{entralities} \gets \Call{LapCentUnweighted}{snapshot}$
        \Else
            \State$\mathcal{C}_{entralities}$, $\mathcal{N}_{um}\mathcal{C}_{entralities} \gets \Call{LapCentWeighted}{snapshot}$
        \EndIf
        
  \EndFor
  
  \EndProcedure
  
\end{algorithmic}
\end{algorithm}

\end{description}

\subsection{Centrality Measures for Evolving Networks}

Due to requirements for size or dynamics of networks, some centrality measures were already adapted to be incremental or dynamic. The motivation for the use of incremental algorithms vary, but, the main one, is to efficiently process large volumes of data that is subject to many (relatively) small changes over time. In this subsection, we briefly state some of these improvements regarding incremental centrality measure algorithms. The authors of these algorithms argument their solutions are faster than the batch versions. Our objective is to achieve similar improvements, for the batch version of the Laplacian Centrality algorithm.

\begin{description}
    \item [\textbf{Betweenness Centrality}:] The Brandes algorithm \cite{Brandes01afaster} (currently widely used), runs in $O(mn+n^2 log n)$ time, where $n = \arrowvert V \arrowvert$ and $m = \arrowvert E \arrowvert$. Nasre et al. \cite{journals/corr/NasrePR13} developed an incremental algorithm to perform Betweenness Centrality (BC) measures in dynamic networks. The BC score of all vertices in G is updated when a new edge is added to G, or the weight of an existing edge is reduced. Their incremental algorithm runs in $O(m'n+n^2)$ time, where $m'$ is bounded by $m\ast = \arrowvert E\ast \arrowvert$, and $E\ast$ is the set of edges that lie on the shortest path in G. The authors explain that, even for a single edge update, their incremental algorithm is the first algorithm that is faster on sparse graphs when compared with recomputing using the well-known static Brandes algorithm. The authors also stress that their algorithm is also likely to be much faster than Brandes on dense graphs since $m\ast$ is often close to linear in $n$. The authors explain that with preliminary experimental results for their basic edge update algorithm on random graphs, generated using the Erdós-Rényi model, they achieve 2 to 15 times speed-up over Brandes’ algorithm for graphs with 256 to 2048 nodes, with the larger speed-ups on dense graphs. Previously to this update of Betweenness Centrality measurements, Kas et al. had already tried to adapt the algorithm for evolving graphs in \cite{6785684}. We consider both approaches relevant for the research that needs to calculate Betweenness Centrality in evolving graphs.\\
    
    \item [\textbf{Closeness Centrality}:] Recently, two significant publications regarding the update of Closeness Centrality in evolving graphs were proposed by Kas et al. and Sariyuce et al. \cite{Kas:2013:ICC:2492517.2500270,Sariyuce2013} . To provide a conceptual overview to the reader, we will focus on Kas et al. work in this paper.  Kas et al. \cite{Kas:2013:ICC:2492517.2500270} developed an incremental Closeness Centrality algorithm for dynamic networks. To compute the closeness values incrementally, for streaming, dynamically changing social networks, all-pairs shortest-paths algorithm proposed by Ramalingam and Reps \cite{Ramalingam:1996:IAG:235067.235070} was extended, such that closeness values are incrementally updated, in line with the changing shortest path distances in the network. As shown in figure \ref{fig:IncCentrality}, the addition of an edge between $X$ and $Y$ nodes would be processed by discovering affected sources, i.e., nodes that are on the path of the new edge, and the affected sinks, i.e., the nodes that are beyond the added connection. Finally, the authors update the Closeness Centrality values just for the affected nodes. The author's argue that, for incremental algorithms, computation times can benefit from early pruning by updating only the affected parts. While the original algorithm for Closeness Centrality can be performed by running an all-pair shortest paths algorithm like Floyd-Warshall \cite{Floyd:1962:A9S:367766.368168}, which results in $O(n^3)$ time complexity, Kas et al. achieve several improvements to their 2-phase algorithm. They argue that the time complexity of Phase-1 to be limited by $O(||A_{ffected}||_2 )$ where the subscript 2 denotes the size of two-hop neighborhood of all affected nodes. The complexity of Phase-2 is dominated by the complexity of priority queue, denoted by $O(||A_{ffected}|| log ||A_{ffected}||)$. The authors' effort results in significant speed-ups over the most commonly used method, on various synthetic and real-life Datasets, suggesting that incremental algorithm design is a fruitful research area for social network analysts. The speed-ups they achieve with this algorithm vary regarding the topology of the network, and for synthetic networks, they conclude that the incremental algorithm is, on average, six times faster than Dijkstra’s algorithm.
    
    \begin{figure}[H]
        \centering
        \includegraphics[scale=0.031]{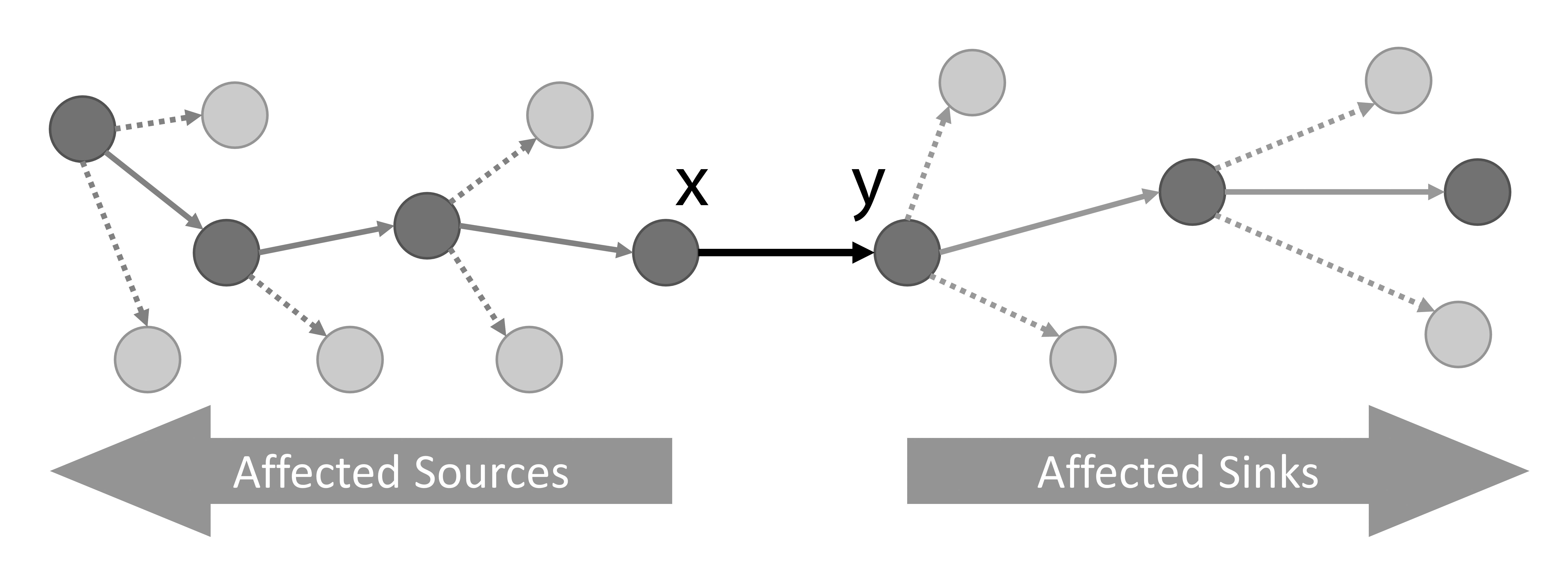}
        \caption{Kas Incremental Closeness Centrality}
        \label{fig:IncCentrality}
    \end{figure}
    
    \item [\textbf{Other Incremental and Dynamic Centrality Measures}:] Other existing centrality measures, for example, Eigenvector centrality, have improvements or variants developed to approach evolving graphs. Eigenvector centrality, or Eigencentrality, has  Google's PageRank and the Katz centrality as possible variants of this measure \cite{AMS}. Regarding these variants, PageRank has been updated to be an incremental algorithm by several researchers, for example, in \cite{Bahmani:2010:FIP:1929861.1929864,Desikan:2005:IPR:1062745.1062885,Kim:2015:IIM:2701126.2701165}. These improvements over the original PageRank measure show significantly faster results when compared with the original PageRank that, for being an iterative process, did not scale well to large-scale graphs.
\end{description}

\section{Proposal for a Dynamic Laplace Centrality algorithm}
\label{IncrementalLaplaceCentralityalgorithm}

\subsection{Locality of the Laplacian Centrality}
\label{LocalityOfTheLaplacianCentrality}

\begin{figure}
  \centering
  \begin{minipage}[b]{.43\columnwidth}
    \subfloat[][Batch Algorithm]{

\resizebox{0.6\linewidth}{!}{
    \begin{tikzpicture}

        \tikzstyle{node} = [fill=black!5]
    
        \vertex[fill=black!50](l1) at (0,0) {1};
        \vertex[fill=black!50](l2) at ([shift={(l1)}] -120:2.33) {2};
        \vertex[fill=black!50](l3) at ([shift={(l2)}] 0:2.33) {3};
        
        \vertex[fill=black!50](l5) at ([shift={(l3)}] 0:2) {5};
        \vertex[fill=black!50](l4) at ([shift={(l5)}] 90:2) {4};
        \vertex[fill=black!50](l6) at ([shift={(l5)}] 0:2) {6};
        \vertex[fill=black!50](l7) at ([shift={(l6)}] 90:2) {7};
        
        \tikzset{EdgeStyle/.style={-, line width=0.3}}
        \Edge[](l1)(l2)
        \Edge[](l1)(l3)
        \Edge[](l2)(l3)
        \Edge[](l3)(l5)
        \Edge[](l5)(l4)
        \Edge[](l5)(l6)
        \Edge[](l6)(l7)
        \Edge[](l4)(l7);
        
        \tikzset{EdgeStyle/.style={dashed, -, line width=0.3}}
        
        \Edge[](l4)(l6)

    \end{tikzpicture}
}
    
    \label{fig:Batch:ToyExample}}
    \end{minipage}%
    \begin{minipage}[b]{.45\columnwidth}
    \subfloat[][Incremental Algorithm]{

\resizebox{0.6\linewidth}{!}{
    \begin{tikzpicture}

        \tikzstyle{node} = [fill=black!5]
    
        \vertex[node](l1) at (0,0) {1};
        \vertex[node](l2) at ([shift={(l1)}] -120:2.33) {2};
        \vertex[node](l3) at ([shift={(l2)}] 0:2.33) {3};
        
        \vertex[fill=black!25](l5) at ([shift={(l3)}] 0:2) {5};
        \vertex[fill=black!50](l4) at ([shift={(l5)}] 90:2) {4};
        \vertex[fill=black!50](l6) at ([shift={(l5)}] 0:2) {6};
        \vertex[fill=black!25](l7) at ([shift={(l6)}] 90:2) {7};
        
        \tikzset{EdgeStyle/.style={-, line width=0.3}}
        \Edge[](l1)(l2)
        \Edge[](l1)(l3)
        \Edge[](l2)(l3)
        \Edge[](l3)(l5)
        \Edge[](l5)(l4)
        \Edge[](l5)(l6)
        \Edge[](l6)(l7)
        \Edge[](l4)(l7);
        
        \tikzset{EdgeStyle/.style={dashed, -, line width=0.3}}
        
        \Edge[](l4)(l6)

    \end{tikzpicture}
}

    \label{fig:Dynamic:ToyExample}}
  \end{minipage}
\caption{Calculated node centralities with edge \{(4, 6)\} added. Dark grey nodes affected by addition of edges. Light grey nodes centralities need to be calculated due to their neighbourhood with affected nodes.}
\label{fig:BatchVsDynamic}
\end{figure}
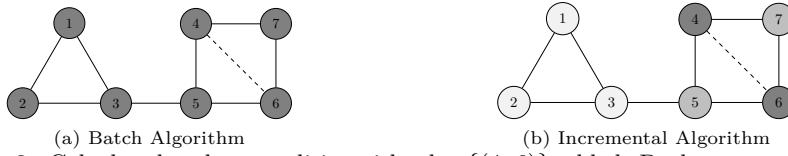

As already stated before, the Laplacian Centrality metric is not a global measure, i.e., is a function of the local degree plus the degree’s of the neighbors (with different weights for each). Xingqin Qi et al. \cite{Qi:2012:LCN:2181343.2181780,Qi2013}, and the pseudo-code presented in \algorithmref{BatchLaplaceCentralityAlgorithm}, shown that local degree and the 1st order neighbors degree is all that is needed to calculate the metric for unweighted networks. To demonstrate this property, the toy network example in \figref{fig:BatchVsDynamic} is used to show the locality of the Laplacian Centrality.

 Due to the determinism of the algorithm, the node centrality results for the incremental Laplace algorithm are equal to the results of the batch version of the same algorithm. This is explained by the fact that we are not dealing with probabilistic phenomena nor any randomness in the initialization. The results of both versions of the algorithm have the same values for each node in the dynamic graph of \figref{fig:BatchVsDynamic}, as we can see in \tableref{Table:ExampleofCentralitycalculation}. The obtained values for centrality are the same for both algorithms (expected), but for step 2 in the incremental algorithm, we only calculated the centralities of the affected nodes plus their neighbors. In total, the Batch algorithm performed 14 node centrality calculations (7 for step 1 and 7 for step 2) while the incremental achieved the same result using only 11 node centrality calculations (7 for step 1 and 4 for step 2).

\begin{table}
\centering
\caption{Example of Centrality calculation for the network presented in \figref{fig:BatchVsDynamic}.}
\label{Table:ExampleofCentralitycalculation}
\begin{adjustbox}{max width=0.85\textwidth}
\begin{tabular}{ccccccccc}
\hline
\multirow{3}{*}{\textbf{Node}}                  & \multicolumn{4}{c}{\textbf{Batch}}                                          & \multicolumn{4}{c}{\textbf{Incremental}}                                        \\ 
                                                & \multicolumn{2}{c}{\textbf{step=1}}  & \multicolumn{2}{c}{\textbf{step=2}}  & \multicolumn{2}{c}{\textbf{step=1}}  & \multicolumn{2}{c}{\textbf{step=2}}  \\ \cline{2-9} 
                                                & Centrality        & Calculated       & Centrality        & Calculated       & Centrality        & Calculated       & Centrality      & Calculated         \\ \hline
1                                               & 6                 & yes              & 6                 & yes              & 6                 & yes              & 6               & no                 \\
2                                               & 12                & yes              & 12                & yes              & 12                & yes              & 12              & no                 \\
3                                               & 18                & yes              & 18                & yes              & 18                & yes              & 18              & no                 \\
4                                               & 18                & yes              & 28                & yes              & 18                & yes              & 28              & yes                \\
5                                               & 34                & yes              & 38                & yes              & 34                & yes              & 38              & yes (neighbour)    \\
6                                               & 10                & yes              & 20                & yes              & 10                & yes              & 20              & yes                \\
7                                               & 18                & yes              & 20                & yes              & 18                & yes              & 20              & yes (neighbour)    \\ \hline
\multicolumn{1}{l}{\textbf{Total:}} & \multicolumn{2}{c}{7 out of 7 nodes} & \multicolumn{2}{c}{7 out of 7 nodes} & \multicolumn{2}{c}{7 out of 7 nodes} & \multicolumn{2}{c}{4 out of 7 nodes} \\ 
\multicolumn{1}{l}{} & \multicolumn{4}{c}{14 centrality calculations}  & \multicolumn{4}{c}{11 centrality calculations} \\ \hline
\end{tabular}
\end{adjustbox}
\end{table}

 In the toy network of \figref{fig:BatchVsDynamic} are considered two distinct snapshots: \\
 \noindent
 $G_0 = \{(1, 2), (2, 3), (3, 5), (5, 6), (5, 4), (4, 7), (5, 7)\}$, and $G_1$, a second snapshot, where a new edge $(4, 6)$ is added to the network, so $G_1 = G_0 \cup \{(4, 6)\}$. In the first snapshot ($G_0$), the algorithm needs to perform calculations of centralities for all the nodes in the network. In the second snapshot ($G_1$), once that only local degree and 1st order neighbors degree will affect the centrality values for nodes 4 and 6, the algorithm will just require to calculate the centralities for nodes 4 and 6 (i.e.: affected nodes) and nodes 5 and 7 (1st order neighbors). Here we can note a substantial difference between the batch and the incremental algorithm. In the former, for the second snapshot ($G_1$), all centralities need to be calculated while for the incremental algorithm, only the affected nodes require the calculation of its centrality.

\subsection{Dynamic Laplace Centrality algorithm}

The original Laplace Centrality algorithm proposed by Xingqin Qi et al. \cite{Qi:2012:LCN:2181343.2181780} was primarily designed for static networks (i.e., networks that do not evolve). Nevertheless, being a static algorithm, it can be used to calculate centralities in changing networks with the penalty of performing a full computation of the centralities in each one of the network snapshots. With our proposal, the same Xingqin Qi et al. \cite{Qi:2012:LCN:2181343.2181780} principles were adapted for two incremental algorithms. The proposed incremental algorithm in \cite{sarmento2017efficient} presents better computational efficiency, by performing selective Laplace Centrality calculations only for the nodes affected by the addition and removal of edges in each one of the snapshots (i.e., it reuses information of the previous snapshot to perform the Laplace Centrality calculations on the current snapshot). This is true for unweighted networks only. Although the algorithm is called incremental, because it avoids full calculations, in each one of the snapshots, it is also prepared for the addition and removal of edges in each one of the increments (i.e., \textit{full dynamic} algorithm).

Based on the assumptions devised in the \sectionref{LocalityOfTheLaplacianCentrality} and having by reference the batch version of the Laplacian centrality shown in \algorithmref{BatchLaplaceCentralityAlgorithm}, a new incremental algorithm was presented (Unweighted version \algorithmref{DynamicLaplaceCentralityAlgorithmUnweighted} and Weighted version \algorithmref{DynamicLaplaceCentralityAlgorithmUnweighted}) \footnote{source code available here: https://github.com/mmfcordeiro/DynamicLaplaceCentrality \label{SrcCode}}. This incremental algorithm achieves better efficiency than the batch version by performing Laplace centrality calculations only in the nodes affected by addition or removal of edges and their 1st order neighbors (full dynamic incremental algorithm). In both versions,  \algorithmref{DynamicLaplaceCentralityAlgorithmUnweighted} \textsuperscript{\ref{SrcCode}} and \algorithmref{DynamicLaplaceCentralityAlgorithmWeighted} \textsuperscript{\ref{SrcCode}}, the incremental method \code{LapCentAddRemove()} receives a full graph (network of the previous snapshot) as parameters. It also receives the lists of edges that will be added and removed from the graph in the current iteration ($A_{dd}$ and $R_{emove}$ respectively), and the previously calculated centralities to be updated in the current iteration ($C_{entralities}$). Previously to the beginning of the \code{for each} cycle, the algorithm will calculate the set of nodes affected by the addition and removal of nodes ($\mathcal{V}_{s}$) and the list of 1st order neighbors of affected nodes ($\mathcal{V}_{f}$). Laplace centrality will only be calculated for this two sets of nodes using the same centrality calculation function employed in the batch algorithm. Function \code{LapCentAddRemove()} returns the updated graph $G$, the updated centrality list $C_{centralities}$, and $\mid\mathcal{V}_{f}\mid$ the number of nodes for which new centralities were calculated for the current iteration. In the main function, initially, in the first iteration, the centralities are calculated for the full network, and this information is reused in the following iterations. In each of the incremental steps, the function \code{LapCentAddRemove()} only receives the list of edges that changed from the previous iteration. By analysis of the proposed algorithm, we can conclude that obtained efficiency can be related to the number of edges that change in each of the snapshots, and also the degree of its nodes. Higher degrees will require performing more computations of 1st order neighbors. Please remark that \code{LapCentWeighted()} is the weighted version of the unweighted \code{LapCent()} while \code{LapCentWeightedAddRemove()} the weighted version of \code{LapCentAddRemove()}.

The complexity of the original algorithm \cite{Qi:2012:LCN:2181343.2181780} is $O(n * \Delta^2)$, where $n$ is the number of vertices and $\Delta$ as the maximum degree ($\Delta = max_{v\in V(G)}d_{v}$). Thus, the total complexity for computing Laplacian centrality for network
$G$ with $n$ vertices, $m$ edges and maximum degree $\Delta$ would be no more than $O(n*\Delta^2)$. Nonetheless, according to \cite{Wheeler2015} it should be something like $O(n*a)$, where $a$ is the average number of neighbors for the entire graph, and $n$ is the number of nodes. In the worst case, $a$ is the maximum number of neighbors any node has in the graph. Our improvement of this algorithm brings the use of locality features of the original algorithm to lower the complexity to $O(n'*a)$, where $n'$ are the affected nodes in each snapshot of the evolving changes of the networks. The affected nodes by addition or removal of $m'$ edges is given by $n' = (2*m' + 2*m'*a)$, i.e.: the sum of 2 nodes per modified edge $(2*m')$ and their respective 1st order neighbours $(2*m'*\Delta)$ or $(2*m'*a)$ with \cite{Wheeler2015} assumption. Final time complexity is $O(2*m'*a + 2*m'*a^2)$.

\begin{algorithm}[!th]
  \small
  \footnotesize
  \caption{Dynamic Laplace Centrality Algorithm (unweighted version)}\label{DynamicLaplaceCentralityAlgorithmUnweighted}
  \begin{algorithmic}[1]
    \State $V\gets \{u_1,u_2, ..,u_v\}$ ,  $E\gets \{(i_1,j_1),(i_2,j_2), ..,(i_e,j_e)\}$
    \State $A_{dd}\gets array\{(i_1, j_1), .., (i_n, j_n)\}$, $R_{emove}\gets array\{(i_1, j_1), .., (i_n, j_n)\}$ 

    \Procedure{LapCentAddRemove}{$G\gets (V,E)$, $A_{dd}$, $R_{emove}$, $\mathcal{C}_{entralities}$}
    
    \State$\mathcal{V}_{s} \gets \{\}$
    
    \For{\textbf{each} $edge$ \textbf{in} $A_{dd}$}
        \State$\mathcal{V}_{s} \gets \mathcal{V}_{s} \cup edge.source() \cup edge.destination()$
        \State $G.add\_edge(edge)$
    \EndFor
    
    \For{\textbf{each} $edge$ \textbf{in} $R_{emove}$}
        \State$\mathcal{V}_{s} \gets \mathcal{V}_{s} \cup edge.source() \cup edge.destination()$
    \EndFor
    
    $\mathcal{V}_{f} \gets \{\}$
    \For{\textbf{each} $node$ \textbf{in} $\mathcal{V}_{s}$}
        \State$\mathcal{V}_{f} \gets \mathcal{V}_{f} \cup G.neighbors(node)$
    \EndFor
    
    \For{\textbf{each} $edge$ \textbf{in} $R_{emove}$}
        \State $G.remove\_edge(edge)$
    \EndFor
    
    \State$\mathcal{D}_{egrees} \gets G.degrees()$
    

    \For{\textbf{each} $v$ \textbf{in} $\mathcal{V}_{f}$}
    
        \State$\mathcal{N}_{eighbors} \gets G.neighbors()$
        \State$loc \gets \mathcal{D}_{egrees}[v]$
        \State$nei \gets 2 . \sum\limits_{i=1}^{\mathcal{N}_{eighbors}} \mathcal{D}_{egrees}[i]$
        \State$\mathcal{C}_{entralities} [v] \gets ({loc}^2 + loc + nei)$
      
    \EndFor
    
    \State\Return $\mathcal{C}_{entralities}$, $\mid\mathcal{V}_{f}\mid$, $G$

  \EndProcedure

  \Procedure{Main}{}
  
  \State $\mathcal{D}_{ataset} \gets \{G_{0}, G_{1}, ... ,G_{n}\}$, $A_{dd}\gets \{A_{0}, A_{1}, ... ,A_{n}\}$, $R_{emove}\gets \{R_{0}, R_{1}, ... ,R_{n}\}$ 
  
  \State$\mathcal{C}_{entralities}$, $\mathcal{N}_{um}\mathcal{C}_{entralities} \gets \Call{LapCent}{G_{0}}$ \Comment{initial step in the full network}

  \State $G \gets G_{0}$, $i \gets 1$
  \While{$(i \le \vert \mathcal{D}_{ataset} \vert)$} \Comment{calculate centralities for the increments}
        \State$\mathcal{C}_{entralities}$, $\mathcal{N}_{um}\mathcal{C}_{entralities}$, $G$ $\gets \Call{LapCentAddRemove}{G, A_{dd}[i], R_{emove}[i], \mathcal{C}_{entralities}}$
  \EndWhile
  
  \EndProcedure
  
\end{algorithmic}
\end{algorithm}

\begin{algorithm}[!th]
  \small
  \footnotesize
  \caption{Dynamic Laplace Centrality Algorithm (weighted version)}\label{DynamicLaplaceCentralityAlgorithmWeighted}
  \begin{algorithmic}[1]
    \State $V\gets \{u_1,u_2, ..,u_v\}$ ,  $E\gets \{(i_1,j_1),(i_2,j_2), ..,(i_e,j_e)\}$
    \State $A_{dd}\gets array\{(i_1, j_1), .., (i_n, j_n)\}$, $R_{emove}\gets array\{(i_1, j_1), .., (i_n, j_n)\}$ 

    \Procedure{LapCentWeightedAddRemove}{$G\gets (V,E)$, $A_{dd}$, $R_{emove}$, $\mathcal{C}_{entralities}$}
    
    \State$\mathcal{V}_{s} \gets \{\}$
    
    \For{\textbf{each} $edge$ \textbf{in} $A_{dd}$}
        \State$\mathcal{V}_{s} \gets \mathcal{V}_{s} \cup edge.source() \cup edge.destination()$
        \State $G.add\_edge(edge)$
    \EndFor
    
    \For{\textbf{each} $edge$ \textbf{in} $R_{emove}$}
        \State$\mathcal{V}_{s} \gets \mathcal{V}_{s} \cup edge.source() \cup edge.destination()$
    \EndFor
    
    $\mathcal{V}_{f} \gets \{\}$
    \For{\textbf{each} $node$ \textbf{in} $\mathcal{V}_{s}$}
        \State$\mathcal{V}_{f} \gets \mathcal{V}_{f} \cup G.neighbors(node)$
    \EndFor
    
    \For{\textbf{each} $edge$ \textbf{in} $R_{emove}$}
        \State $G.remove\_edge(edge)$
    \EndFor
    
    \State$\mathcal{D}_{egrees} \gets G.degrees()$
    

    \For{\textbf{each} $v$ \textbf{in} $\mathcal{V}_{f}$}
    
        \State$loc \gets \mathcal{D}_{egrees}[v]$
        \State$cw$, $sub$ $\gets \Call{CW}{G, v, \mathcal{D}_{egrees}}$ \Comment{calls centrality weight for weighted networks}
        
        \State$\mathcal{C}_{entralities} [v] \gets ({loc}^2 - sub + 2*cw)$
      
    \EndFor
    
    \State\Return $\mathcal{C}_{entralities}$, $\mid\mathcal{V}_{f}\mid$, $G$

  \EndProcedure

  \Procedure{Main}{}
  
  \State $\mathcal{D}_{ataset} \gets \{G_{0}, G_{1}, ... ,G_{n}\}$, $A_{dd}\gets \{A_{0}, A_{1}, ... ,A_{n}\}$, $R_{emove}\gets \{R_{0}, R_{1}, ... ,R_{n}\}$ 
  
  \State$\mathcal{C}_{entralities}$, $\mathcal{N}_{um}\mathcal{C}_{entralities} \gets \Call{LapCentWeighted}{G_{0}}$ \Comment{initial step in the full network}

  \State $G \gets G_{0}$, $i \gets 1$
  \While{$(i \le \vert \mathcal{D}_{ataset} \vert)$} \Comment{calculate centralities for the increments}
        \State$\mathcal{C}_{entralities}$, $\mathcal{N}_{um}\mathcal{C}_{entralities}$, $G$ $\gets \Call{LapCentWeightedAddRemove}{G, A_{dd}[i], R_{emove}[i], \mathcal{C}_{entralities}}$
  \EndWhile
  
  \EndProcedure
  
\end{algorithmic}
\end{algorithm}

\section{Experimental Results}\label{Res}

The experimental results were obtained using different size networks. The High-energy physics theory citation network \cite{Leskovec2005} has 27 770 vertices and 352 807 edges, the Autonomous systems AS-733 dataset \cite{Leskovec2005} has 6 474 vertices and 13 895 edges in 733 daily snapshots, and the AS Caida Relationships Datasets \cite{Leskovec2005} has 26 475 vertices and 106 762 edges. The Bitcoin Alpha trust weighted signed network has 3 783 vertices and 24 186 edges \cite{kumar2016edge}, the Reuters terror news network (DaysAll) \cite{Batagelj03densitybased} has 13 308 vertices and 148 035 edges distributed by 66 snapshots (1 per day).

The performed empirical evaluation consisted mainly in comparing run times of each increment (duration of each increment and cumulative execution time). Additionally, the size of the network (number of nodes and edges), the number of added/removed edges in each snapshot (negative values mean more edges were removed than added), and the total number of calculated centralities for each snap shot was also registered. The reader should note that the batch algorithm will always need to calculate centralities for all nodes in the snapshot while it is expected that the Dynamic algorithm only performs centrality calculations for the affected nodes. In the end, an analysis of the total speed-up ratio obtained in each of the steps is provided. For all the experimentation and development, we used an Intel (R) Core (TM) i7-4702MQ processor computer with 8 GBytes and SSD HDD. Three runs per algorithm/ dataset were performed with values presented in graphs as the average values.

\subsection{Evaluation Setups}

The Dynamic Laplace Centrality algorithm was evaluated in incremental and dynamic network setups, results are presented in \sectionref{TestScenario:IncrementalNetworkSetup} and \sectionref{TestScenario:DynamicNetworkSetup} respectively. The algorithms were tested in their unweighted and weighted variants in each one of those network scenarios:

\begin{description}
    \item [\textbf{Incremental Networks Evaluation}:] The datasets used for the incremental evaluation were the the High-energy physics theory citation network \cite{Leskovec2005} for unweighted networks tests, and the Reuters terror news network (DaysAll) \cite{Batagelj03densitybased} for weighted networks tests. The evaluation of the incremental setup was done using the original Laplace Centrality (now on called Batch) and the proposed Dynamic Laplace Centrality (now on called Dynamic) in an incremental setting configuration. In both cases the original Laplace Centrality served as a baseline. In the incremental setup, we considered all the snapshots of the datasets. In the High-energy physics theory citation network, 136 snapshots were built by aggregating timestamps of citations in a monthly basis. In the Reuters terror news network (DaysAll), 66 snapshots were considered with data but aggregated in a daily basis. Regarding the testing, while in the Batch setup, for every snapshot, the centralities were calculated having the full network as input, for the incremental algorithm, in the first snapshot the full network is passed as input, and in the following snapshots, the algorithm only receives the set of edges added to the network in that snapshot (incremental). \\
    
    \item [\textbf{Dynamic Networks Evaluation}:] The evaluation of the Dynamic Laplace Centrality algorithm was performed in a dynamic network setup (addition and removal of edges between snapshots) for unweighted and weighted networks. The Autonomous systems AS-733 dataset \cite{Leskovec2005} with 733 snapshots from November 8 1997 to January 2 2000 and AS Caida Relationships Datasets \cite{Leskovec2005} with a total of 122 snapshots from January 2004 to November 2007 were used to evaluate unweighted dynamic networks. The evaluation of the Dynamic Laplace Centrality algorithm in a weighted dynamic network setup was performed using the Reuters terror news network (DaysAll) \cite{Batagelj03densitybased} in an 30 days sliding window (addition and removal of edges), and the Bitcoin Alpha trust weighted signed network (Bitcoin-alpha) \cite{kumar2016edge} in two sliding window setups (aggregation by day, 30 day sliding window in the Bitcoin-alpha-day and aggregation by month, 12 month sliding window in the Bitcoin-alpha). 
\end{description}

\subsection{Incremental Network Results}
\label{TestScenario:IncrementalNetworkSetup}

In this subsection, we introduce the reader to the results obtained by the incremental setup of the algorithm for unweighted and weighted networks.

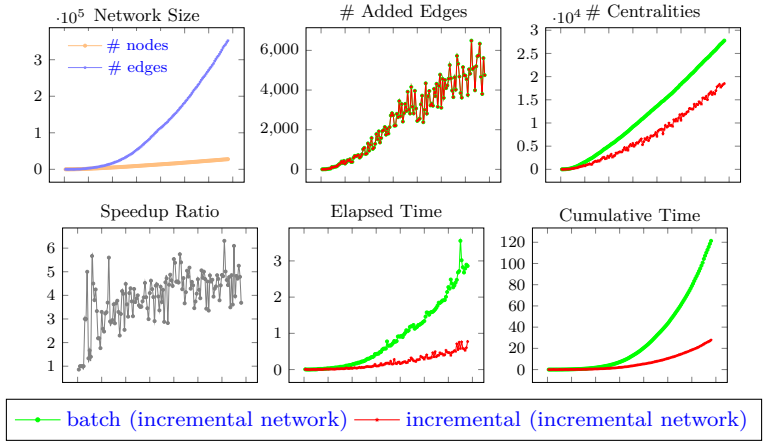
\begin{figure}
  \centering

\pgfplotstableread[col sep=semicolon]{step;number_of_nodes;number_of_edges
1;4;2
2;9;6
3;19;13
4;29;20
5;58;41
6;88;68
7;130;102
8;173;146
9;262;262
10;329;363
11;428;492
12;522;625
13;620;777
14;721;976
15;872;1233
16;1003;1529
17;1155;1919
18;1281;2231
19;1438;2658
20;1555;2948
21;1683;3305
22;1866;3767
23;2015;4203
24;2184;4710
25;2317;5084
26;2445;5460
27;2627;6044
28;2816;6723
29;3008;7388
30;3189;8079
31;3372;8799
32;3518;9390
33;3691;10038
34;3901;10796
35;4115;11806
36;4340;12892
37;4478;13667
38;4645;14466
39;4858;15773
40;5018;16715
41;5212;17950
42;5432;19529
43;5582;20561
44;5754;21659
45;5928;22945
46;6159;24720
47;6377;26234
48;6583;28151
49;6770;29507
50;6956;31087
51;7157;33016
52;7365;34999
53;7604;37131
54;7797;38878
55;8011;40924
56;8211;42728
57;8445;44859
58;8695;47577
59;8934;50410
60;9186;53135
61;9377;55337
62;9577;57546
63;9797;60346
64;10005;62975
65;10253;66417
66;10473;69102
67;10730;72391
68;10902;74799
69;11125;77629
70;11374;80927
71;11583;83869
72;11843;87782
73;12047;90803
74;12255;93617
75;12535;97770
76;12738;100927
77;12942;103746
78;13187;107705
79;13421;110774
80;13608;113216
81;13824;115712
82;14067;118271
83;14320;121545
84;14572;125260
85;14723;127635
86;14916;130633
87;15190;134967
88;15411;137775
89;15635;141453
90;15871;145397
91;16089;148612
92;16296;151925
93;16531;155127
94;16797;159178
95;17040;162864
96;17333;167166
97;17555;170315
98;17803;174525
99;18165;179296
100;18361;182397
101;18645;186946
102;18883;190541
103;19111;195043
104;19348;199133
105;19579;203360
106;19855;207926
107;20149;213176
108;20420;217760
109;20655;221734
110;20854;225368
111;21128;230004
112;21396;234013
113;21717;239726
114;21997;245033
115;22259;249700
116;22512;253544
117;22752;257157
118;23039;262304
119;23321;267046
120;23564;271752
121;23820;276295
122;24023;280041
123;24281;285078
124;24526;289895
125;24844;296379
126;25085;301422
127;25364;306549
128;25583;310525
129;25837;315719
130;26146;321422
131;26432;327167
132;26765;333501
133;27019;338168
134;27241;341968
135;27507;347577
136;27770;352324
}\dataTableNetworksize
\pgfplotstableread[col sep=semicolon]{step;batch;dynamic
1;2;2
2;4;4
3;7;7
4;7;7
5;21;21
6;27;27
7;35;35
8;44;44
9;117;117
10;101;101
11;130;130
12;133;133
13;153;153
14;199;199
15;257;257
16;296;296
17;396;392
18;316;316
19;429;428
20;291;290
21;357;357
22;463;462
23;438;438
24;508;508
25;374;374
26;376;376
27;584;584
28;680;679
29;666;665
30;691;691
31;720;720
32;591;591
33;648;648
34;758;758
35;1012;1011
36;1086;1086
37;776;775
38;802;801
39;1308;1308
40;942;942
41;1235;1235
42;1580;1580
43;1032;1032
44;1099;1098
45;1286;1286
46;1777;1776
47;1514;1514
48;1917;1917
49;1356;1356
50;1580;1580
51;1929;1929
52;1983;1983
53;2135;2134
54;1748;1748
55;2047;2046
56;1804;1804
57;2132;2131
58;2718;2718
59;2833;2833
60;2733;2725
61;2202;2202
62;2211;2210
63;2800;2800
64;2630;2630
65;3442;3442
66;2685;2685
67;3292;3289
68;2409;2408
69;2830;2830
70;3301;3300
71;2943;2942
72;3915;3914
73;3021;3021
74;2814;2814
75;4156;4153
76;3158;3158
77;2821;2821
78;3961;3960
79;3072;3070
80;2442;2442
81;2502;2498
82;2562;2561
83;3275;3274
84;3719;3718
85;2377;2376
86;3003;3000
87;4340;4338
88;2813;2812
89;3689;3685
90;3951;3951
91;3226;3223
92;3317;3315
93;3208;3206
94;4062;4060
95;3692;3690
96;4312;4308
97;3155;3154
98;4218;4216
99;4778;4777
100;3108;3105
101;4568;4554
102;3604;3600
103;4506;4504
104;4097;4094
105;4233;4231
106;4572;4571
107;5261;5255
108;4594;4589
109;3976;3975
110;3642;3639
111;4640;4638
112;4023;4015
113;5724;5720
114;5328;5321
115;4679;4675
116;3852;3848
117;3618;3617
118;5154;5150
119;4753;4748
120;4726;4717
121;4549;4547
122;3751;3749
123;5045;5043
124;4825;4823
125;6494;6490
126;5059;5056
127;5133;5130
128;3986;3984
129;5196;5195
130;5705;5703
131;5747;5745
132;6340;6335
133;4668;4667
134;3800;3800
135;5612;5609
136;4752;4749
}\dataTableAddedEdges
\pgfplotstableread[col sep=semicolon]{step;batch;dynamic
1;4;4
2;9;7
3;19;17
4;29;16
5;58;46
6;88;58
7;130;76
8;173;87
9;262;157
10;329;183
11;428;246
12;522;276
13;620;323
14;721;334
15;872;495
16;1003;574
17;1155;644
18;1281;660
19;1438;830
20;1555;745
21;1683;890
22;1866;1017
23;2015;1064
24;2184;1195
25;2317;1083
26;2445;1092
27;2627;1470
28;2816;1660
29;3008;1740
30;3189;1684
31;3372;1896
32;3518;1748
33;3691;1983
34;3901;2260
35;4115;2539
36;4340;2595
37;4478;2279
38;4645;2351
39;4858;2840
40;5018;2558
41;5212;2951
42;5432;3249
43;5582;2624
44;5754;2848
45;5928;3007
46;6159;3507
47;6377;3458
48;6583;3657
49;6770;3775
50;6956;3757
51;7157;3771
52;7365;4216
53;7604;4702
54;7797;4176
55;8011;4670
56;8211;4115
57;8445;4976
58;8695;5366
59;8934;5081
60;9186;5594
61;9377;4970
62;9577;5263
63;9797;5723
64;10005;5629
65;10253;5942
66;10473;6249
67;10730;6572
68;10902;5296
69;11125;5962
70;11374;6644
71;11583;6423
72;11843;6906
73;12047;6185
74;12255;6759
75;12535;7250
76;12738;7161
77;12942;7234
78;13187;7458
79;13421;7944
80;13608;7117
81;13824;8243
82;14067;8508
83;14320;8600
84;14572;8674
85;14723;7634
86;14916;8618
87;15190;9495
88;15411;9082
89;15635;9263
90;15871;9967
91;16089;9461
92;16296;9512
93;16531;9285
94;16797;10696
95;17040;10389
96;17333;10838
97;17555;10411
98;17803;11057
99;18165;11794
100;18361;10584
101;18645;11959
102;18883;11542
103;19111;11870
104;19348;11995
105;19579;12337
106;19855;12484
107;20149;13126
108;20420;13070
109;20655;13003
110;20854;12997
111;21128;14035
112;21396;13499
113;21717;14033
114;21997;13958
115;22259;14603
116;22512;14335
117;22752;14068
118;23039;15375
119;23321;15398
120;23564;15245
121;23820;15752
122;24023;15023
123;24281;16173
124;24526;16412
125;24844;16764
126;25085;16136
127;25364;16549
128;25583;16255
129;25837;16969
130;26146;17799
131;26432;18049
132;26765;18330
133;27019;17722
134;27241;18032
135;27507;18313
136;27770;18508
}\dataTableNumberCentralities
\pgfplotstableread[col sep=semicolon]{step;batch;dynamic
1;0.006;0.007
2;0.001;0.001
3;0.001;0.000
4;0.001;0.001
5;0.001;0.001
6;0.003;0.001
7;0.003;0.001
8;0.005;0.001
9;0.004;0.003
10;0.005;0.003
11;0.007;0.005
12;0.017;0.003
13;0.018;0.004
14;0.019;0.005
15;0.017;0.004
16;0.020;0.006
17;0.024;0.011
18;0.024;0.011
19;0.030;0.017
20;0.033;0.010
21;0.031;0.012
22;0.037;0.015
23;0.042;0.013
24;0.046;0.014
25;0.048;0.013
26;0.056;0.010
27;0.055;0.019
28;0.060;0.021
29;0.062;0.021
30;0.076;0.021
31;0.076;0.027
32;0.077;0.021
33;0.083;0.022
34;0.089;0.027
35;0.108;0.047
36;0.106;0.033
37;0.117;0.028
38;0.122;0.030
39;0.127;0.050
40;0.139;0.031
41;0.152;0.042
42;0.158;0.051
43;0.172;0.042
44;0.172;0.040
45;0.191;0.051
46;0.190;0.061
47;0.205;0.048
48;0.223;0.060
49;0.224;0.058
50;0.236;0.059
51;0.254;0.071
52;0.281;0.080
53;0.299;0.085
54;0.300;0.080
55;0.317;0.081
56;0.344;0.076
57;0.345;0.088
58;0.366;0.106
59;0.428;0.143
60;0.422;0.103
61;0.439;0.095
62;0.439;0.103
63;0.483;0.123
64;0.486;0.167
65;0.547;0.124
66;0.575;0.129
67;0.581;0.147
68;0.579;0.172
69;0.777;0.178
70;0.658;0.177
71;0.665;0.157
72;0.662;0.230
73;0.692;0.152
74;0.732;0.155
75;0.789;0.279
76;0.766;0.172
77;0.785;0.161
78;0.913;0.197
79;0.915;0.208
80;0.908;0.164
81;0.931;0.173
82;0.897;0.196
83;0.964;0.210
84;1.009;0.222
85;1.085;0.189
86;1.058;0.196
87;1.082;0.254
88;1.178;0.249
89;1.194;0.326
90;1.195;0.289
91;1.231;0.244
92;1.127;0.235
93;1.159;0.224
94;1.178;0.275
95;1.230;0.269
96;1.234;0.279
97;1.262;0.365
98;1.279;0.342
99;1.315;0.323
100;1.339;0.286
101;1.461;0.345
102;1.578;0.407
103;1.568;0.310
104;1.514;0.319
105;1.571;0.314
106;1.591;0.339
107;1.594;0.464
108;1.624;0.354
109;1.661;0.495
110;1.705;0.352
111;1.765;0.385
112;1.846;0.396
113;1.839;0.466
114;1.867;0.436
115;1.905;0.435
116;1.912;0.394
117;1.920;0.409
118;2.009;0.530
119;2.141;0.484
120;2.276;0.468
121;2.233;0.464
122;2.486;0.394
123;2.375;0.474
124;2.270;0.488
125;2.323;0.524
126;2.401;0.503
127;2.477;0.709
128;2.680;0.552
129;2.718;0.753
130;3.554;0.583
131;3.018;0.762
132;2.826;0.586
133;2.679;0.569
134;2.790;0.531
135;2.895;0.605
136;2.860;0.776
}\dataTableElapsedTime
\pgfplotstableread[col sep=semicolon]{step;batch;dynamic
1;0.006;0.007
2;0.007;0.008
3;0.008;0.008
4;0.009;0.009
5;0.01;0.01
6;0.013;0.011
7;0.016;0.012
8;0.021;0.013
9;0.025;0.016
10;0.03;0.019
11;0.037;0.024
12;0.054;0.027
13;0.072;0.031
14;0.091;0.036
15;0.108;0.04
16;0.128;0.046
17;0.152;0.057
18;0.176;0.068
19;0.206;0.085
20;0.239;0.095
21;0.27;0.107
22;0.307;0.122
23;0.349;0.135
24;0.395;0.149
25;0.443;0.162
26;0.499;0.172
27;0.554;0.191
28;0.614;0.212
29;0.676;0.233
30;0.752;0.254
31;0.828;0.281
32;0.905;0.302
33;0.988;0.324
34;1.077;0.351
35;1.185;0.398
36;1.291;0.431
37;1.408;0.459
38;1.53;0.489
39;1.657;0.539
40;1.796;0.57
41;1.948;0.612
42;2.106;0.663
43;2.278;0.705
44;2.45;0.745
45;2.641;0.796
46;2.831;0.857
47;3.036;0.905
48;3.259;0.965
49;3.483;1.023
50;3.719;1.082
51;3.973;1.153
52;4.254;1.233
53;4.553;1.318
54;4.853;1.398
55;5.17;1.479
56;5.514;1.555
57;5.859;1.643
58;6.225;1.749
59;6.653;1.892
60;7.075;1.995
61;7.514;2.09
62;7.953;2.193
63;8.436;2.316
64;8.922;2.483
65;9.469;2.607
66;10.044;2.736
67;10.625;2.883
68;11.204;3.055
69;11.981;3.233
70;12.639;3.41
71;13.304;3.567
72;13.966;3.797
73;14.658;3.949
74;15.39;4.104
75;16.179;4.383
76;16.945;4.555
77;17.73;4.716
78;18.643;4.913
79;19.558;5.121
80;20.466;5.285
81;21.397;5.458
82;22.294;5.654
83;23.258;5.864
84;24.267;6.086
85;25.352;6.275
86;26.41;6.471
87;27.492;6.725
88;28.67;6.974
89;29.864;7.3
90;31.059;7.589
91;32.29;7.833
92;33.417;8.068
93;34.576;8.292
94;35.754;8.567
95;36.984;8.836
96;38.218;9.115
97;39.48;9.48
98;40.759;9.822
99;42.074;10.145
100;43.413;10.431
101;44.874;10.776
102;46.452;11.183
103;48.02;11.493
104;49.534;11.812
105;51.105;12.126
106;52.696;12.465
107;54.29;12.929
108;55.914;13.283
109;57.575;13.778
110;59.28;14.13
111;61.045;14.515
112;62.891;14.911
113;64.73;15.377
114;66.597;15.813
115;68.502;16.248
116;70.414;16.642
117;72.334;17.051
118;74.343;17.581
119;76.484;18.065
120;78.76;18.533
121;80.993;18.997
122;83.479;19.391
123;85.854;19.865
124;88.124;20.353
125;90.447;20.877
126;92.848;21.38
127;95.325;22.089
128;98.005;22.641
129;100.723;23.394
130;104.277;23.977
131;107.295;24.739
132;110.121;25.325
133;112.8;25.894
134;115.59;26.425
135;118.485;27.03
136;121.345;27.806
}\dataTableCumulativeTime
\pgfplotstableread[col sep=semicolon]{step;speedup
1;0.857143
2;1
3;1
4;1
5;1
6;3
7;3
8;5
9;1.33333
10;1.66667
11;1.4
12;5.66667
13;4.5
14;3.8
15;4.25
16;3.33333
17;2.18182
18;2.18182
19;1.76471
20;3.3
21;2.58333
22;2.46667
23;3.23077
24;3.28571
25;3.69231
26;5.6
27;2.89474
28;2.85714
29;2.95238
30;3.61905
31;2.81481
32;3.66667
33;3.77273
34;3.2963
35;2.29787
36;3.21212
37;4.17857
38;4.06667
39;2.54
40;4.48387
41;3.61905
42;3.09804
43;4.09524
44;4.3
45;3.7451
46;3.11475
47;4.27083
48;3.71667
49;3.86207
50;4
51;3.57746
52;3.5125
53;3.51765
54;3.75
55;3.91358
56;4.52632
57;3.92045
58;3.45283
59;2.99301
60;4.09709
61;4.62105
62;4.26214
63;3.92683
64;2.91018
65;4.41129
66;4.45736
67;3.95238
68;3.36628
69;4.36517
70;3.71751
71;4.23567
72;2.87826
73;4.55263
74;4.72258
75;2.82796
76;4.45349
77;4.87578
78;4.63452
79;4.39904
80;5.53659
81;5.3815
82;4.57653
83;4.59048
84;4.54505
85;5.74074
86;5.39796
87;4.25984
88;4.73092
89;3.66258
90;4.13495
91;5.04508
92;4.79574
93;5.17411
94;4.28364
95;4.57249
96;4.42294
97;3.45753
98;3.73977
99;4.07121
100;4.68182
101;4.23478
102;3.87715
103;5.05806
104;4.74608
105;5.00318
106;4.69322
107;3.43534
108;4.58757
109;3.35556
110;4.84375
111;4.58442
112;4.66162
113;3.94635
114;4.28211
115;4.37931
116;4.85279
117;4.69438
118;3.79057
119;4.42355
120;4.86325
121;4.8125
122;6.30964
123;5.01055
124;4.65164
125;4.43321
126;4.77336
127;3.49365
128;4.85507
129;3.60956
130;6.09605
131;3.96063
132;4.82253
133;4.70826
134;5.25424
135;4.78512
136;3.68557
}\dataTableSpeedup

\pgfplotsset{footnotesize}
\begin{center}
%
%
\begin{tikzpicture}[scale=0.75]
\begin{axis}[
title={Network Size},
xticklabels={,,},
legend style={draw=none,legend to name=leg}
]
\addplot[smooth,mark=*,orange!50,mark size=0.75] table[y = number_of_nodes] from \dataTableNetworksize;
\addplot[smooth,mark=star,blue!50,mark size=0.75] table[y = number_of_edges] from \dataTableNetworksize;

   \addlegendentry{\# nodes}
   \addlegendentry{\# edges}

   \node [above left] (L) at (rel axis cs:0.7,0.6) {\ref{leg}};

\end{axis}
\end{tikzpicture}
\begin{tikzpicture}[scale=0.75]
\begin{axis}[
title={\# Added Edges},
xticklabels={,,},
]
\addplot[smooth,mark=*,green,mark size=0.75] table[y = batch] from \dataTableAddedEdges;
\addplot[smooth,mark=star,red,mark size=0.75] table[y = dynamic] from \dataTableAddedEdges;
\end{axis}
\end{tikzpicture}
\begin{tikzpicture}[scale=0.75]
\begin{axis}[
title={\# Centralities},
xticklabels={,,},
]
\addplot[smooth,mark=*,green,mark size=0.75] table[y = batch] from \dataTableNumberCentralities;
\addplot[smooth,mark=star,red,mark size=0.75] table[y = dynamic] from \dataTableNumberCentralities;
\end{axis}
\end{tikzpicture}
\\
\begin{tikzpicture}[scale=0.75]
\begin{axis}[
title={Speedup Ratio},
xticklabels={,,},
legend style={draw=none}
]
\addplot[smooth,mark=*,black!50,mark size=0.75] table[y = speedup] from \dataTableSpeedup;

\end{axis}
\end{tikzpicture}
\begin{tikzpicture}[scale=0.75]
\begin{axis}[
title={Elapsed Time},
xticklabels={,,},
]
\addplot[smooth,mark=*,green,mark size=0.75] table[y = batch] from \dataTableElapsedTime;
\addplot[smooth,mark=star,red,mark size=0.75] table[y = dynamic] from \dataTableElapsedTime;
\end{axis}
\end{tikzpicture}
\begin{tikzpicture}[scale=0.75]
\begin{axis}[
legend columns=-1,
legend entries={batch (incremental network), incremental (incremental network)},
legend to name=named,
title={Cumulative Time},
xticklabels={,,},
]
\addplot[smooth,mark=*,green,mark size=0.75] table[y = batch] from \dataTableCumulativeTime;
\addplot[smooth,mark=star,red,mark size=0.75] table[y = dynamic] from \dataTableCumulativeTime;
\end{axis}
\end{tikzpicture}
\\
\ref{named}
\end{center}
    \caption{Results for the Cit-HepTh (incremental only, unweighted)}
\label{fig:Results1}
\end{figure}

\begin{figure}[!th]
  \centering

\pgfplotstableread[col sep=semicolon]{step;number_of_nodes;number_of_edges
1;2420;9318
2;4169;21396
3;4755;26943
4;5249;31337
5;5699;35883
6;6210;41402
7;6513;44616
8;6706;46930
9;6741;47283
10;7020;50563
11;7310;53705
12;7448;55468
13;7529;56646
14;7641;58202
15;7967;62137
16;8263;65419
17;8473;67811
18;8623;69620
19;8699;70658
20;8808;72342
21;8945;74551
22;9078;76621
23;9193;78405
24;9305;79884
25;9493;82450
26;9583;84133
27;9737;86310
28;9861;88330
29;9975;90006
30;10101;91803
31;10281;94307
32;10416;96227
33;10535;97946
34;10617;99466
35;10713;101443
36;10887;103739
37;11009;105882
38;11099;107474
39;11232;109483
40;11311;111191
41;11374;112564
42;11434;113707
43;11505;114839
44;11603;116476
45;11714;117903
46;11789;119224
47;11865;120735
48;11961;122265
49;12038;123908
50;12155;125722
51;12244;127281
52;12354;129227
53;12440;131073
54;12561;132861
55;12616;134088
56;12742;136189
57;12799;137205
58;12883;138477
59;12959;140057
60;13032;141547
61;13105;142897
62;13145;143858
63;13197;144891
64;13240;146015
65;13265;146984
66;13308;148035
}\dataTableNetworksize
\pgfplotstableread[col sep=semicolon]{step;batch;dynamic
1;9318;9318
2;12078;12078
3;5547;5547
4;4394;4394
5;4546;4546
6;5519;5519
7;3214;3214
8;2314;2314
9;353;353
10;3280;3280
11;3142;3142
12;1763;1763
13;1178;1178
14;1556;1556
15;3935;3935
16;3282;3282
17;2392;2392
18;1809;1809
19;1038;1038
20;1684;1684
21;2209;2209
22;2070;2070
23;1784;1784
24;1479;1479
25;2566;2566
26;1683;1683
27;2177;2177
28;2020;2020
29;1676;1676
30;1797;1797
31;2504;2504
32;1920;1920
33;1719;1719
34;1520;1520
35;1977;1977
36;2296;2296
37;2143;2143
38;1592;1592
39;2009;2009
40;1708;1708
41;1373;1373
42;1143;1143
43;1132;1132
44;1637;1637
45;1427;1427
46;1321;1321
47;1511;1511
48;1530;1530
49;1643;1643
50;1814;1814
51;1559;1559
52;1946;1946
53;1846;1846
54;1788;1788
55;1227;1227
56;2101;2101
57;1016;1016
58;1272;1272
59;1580;1580
60;1490;1490
61;1350;1350
62;961;961
63;1033;1033
64;1124;1124
65;969;969
66;1051;1051
}\dataTableAddedEdges
\pgfplotstableread[col sep=semicolon]{step;batch;dynamic
1;2420;2420
2;4169;4115
3;4755;4517
4;5249;4892
5;5699;5293
6;6210;5825
7;6513;5901
8;6706;5803
9;6741;4114
10;7020;6415
11;7310;6656
12;7448;6303
13;7529;6124
14;7641;6512
15;7967;7229
16;8263;7516
17;8473;7590
18;8623;7436
19;8699;6901
20;8808;7378
21;8945;7748
22;9078;7872
23;9193;7709
24;9305;7462
25;9493;8332
26;9583;8134
27;9737;8259
28;9861;8685
29;9975;8430
30;10101;8547
31;10281;9222
32;10416;8785
33;10535;9050
34;10617;9008
35;10713;9286
36;10887;9448
37;11009;9548
38;11099;9328
39;11232;9794
40;11311;9568
41;11374;9495
42;11434;9163
43;11505;9385
44;11603;9813
45;11714;9708
46;11789;9655
47;11865;10022
48;11961;10017
49;12038;10278
50;12155;10317
51;12244;10236
52;12354;10642
53;12440;10717
54;12561;10752
55;12616;10266
56;12742;11148
57;12799;9995
58;12883;10279
59;12959;10987
60;13032;11136
61;13105;10760
62;13145;10321
63;13197;10485
64;13240;10453
65;13265;10220
66;13308;10635
}\dataTableNumberCentralities
\pgfplotstableread[col sep=semicolon]{step;batch;dynamic
1;0.127;0.125
2;0.285;0.286
3;0.353;0.187
4;0.412;0.186
5;0.502;0.212
6;0.585;0.277
7;0.604;0.213
8;0.639;0.204
9;0.647;0.155
10;0.679;0.236
11;0.721;0.248
12;0.743;0.240
13;0.760;0.224
14;0.788;0.243
15;0.833;0.292
16;0.879;0.328
17;0.904;0.306
18;0.934;0.278
19;0.944;0.266
20;0.977;0.288
21;0.994;0.302
22;1.036;0.313
23;1.034;0.307
24;1.068;0.306
25;1.090;0.342
26;1.115;0.331
27;1.148;0.400
28;1.176;0.355
29;1.190;0.352
30;1.217;0.359
31;1.258;0.407
32;1.273;0.384
33;1.299;0.382
34;1.326;0.387
35;1.338;0.407
36;1.387;0.410
37;1.385;0.420
38;1.415;0.413
39;1.456;0.477
40;1.461;0.421
41;1.514;0.426
42;1.545;0.437
43;1.577;0.442
44;1.570;0.460
45;1.626;0.456
46;1.649;0.464
47;1.659;0.461
48;1.684;0.460
49;1.669;0.508
50;1.695;0.502
51;1.751;0.483
52;1.764;0.503
53;1.766;0.564
54;1.842;0.505
55;1.833;0.500
56;1.827;0.536
57;1.875;0.527
58;1.867;0.513
59;1.893;0.532
60;1.897;0.537
61;1.940;0.525
62;1.936;0.517
63;1.944;0.525
64;1.963;0.528
65;1.960;0.522
66;1.971;0.550
}\dataTableElapsedTime
\pgfplotstableread[col sep=semicolon]{step;batch;dynamic
1;0.127;0.125
2;0.412;0.411
3;0.765;0.598
4;1.177;0.784
5;1.679;0.996
6;2.264;1.273
7;2.868;1.486
8;3.507;1.69
9;4.154;1.845
10;4.833;2.081
11;5.554;2.329
12;6.297;2.569
13;7.057;2.793
14;7.845;3.036
15;8.678;3.328
16;9.557;3.656
17;10.461;3.962
18;11.395;4.24
19;12.339;4.506
20;13.316;4.794
21;14.31;5.096
22;15.346;5.409
23;16.38;5.716
24;17.448;6.022
25;18.538;6.364
26;19.653;6.695
27;20.801;7.095
28;21.977;7.45
29;23.167;7.802
30;24.384;8.161
31;25.642;8.568
32;26.915;8.952
33;28.214;9.334
34;29.54;9.721
35;30.878;10.128
36;32.265;10.538
37;33.65;10.958
38;35.065;11.371
39;36.521;11.848
40;37.982;12.269
41;39.496;12.695
42;41.041;13.132
43;42.618;13.574
44;44.188;14.034
45;45.814;14.49
46;47.463;14.954
47;49.122;15.415
48;50.806;15.875
49;52.475;16.383
50;54.17;16.885
51;55.921;17.368
52;57.685;17.871
53;59.451;18.435
54;61.293;18.94
55;63.126;19.44
56;64.953;19.976
57;66.828;20.503
58;68.695;21.016
59;70.588;21.548
60;72.485;22.085
61;74.425;22.61
62;76.361;23.127
63;78.305;23.652
64;80.268;24.18
65;82.228;24.702
66;84.199;25.252
}\dataTableCumulativeTime
\pgfplotstableread[col sep=semicolon]{step;speedup
1;1.016
2;0.996503
3;1.8877
4;2.21505
5;2.36792
6;2.11191
7;2.83568
8;3.13235
9;4.17419
10;2.87712
11;2.90726
12;3.09583
13;3.39286
14;3.2428
15;2.85274
16;2.67988
17;2.95425
18;3.35971
19;3.54887
20;3.39236
21;3.29139
22;3.3099
23;3.36808
24;3.4902
25;3.18713
26;3.36858
27;2.87
28;3.31268
29;3.38068
30;3.38997
31;3.09091
32;3.3151
33;3.40052
34;3.42636
35;3.28747
36;3.38293
37;3.29762
38;3.42615
39;3.05241
40;3.47031
41;3.55399
42;3.53547
43;3.56787
44;3.41304
45;3.56579
46;3.55388
47;3.5987
48;3.66087
49;3.28543
50;3.37649
51;3.62526
52;3.50696
53;3.13121
54;3.64752
55;3.666
56;3.40858
57;3.55787
58;3.63938
59;3.55827
60;3.53259
61;3.69524
62;3.74468
63;3.70286
64;3.7178
65;3.75479
66;3.58364
}\dataTableSpeedup

\pgfplotsset{footnotesize}
\begin{center}
%
%
\begin{tikzpicture}[scale=0.75]
\begin{axis}[
title={Network Size},
xticklabels={,,},
legend style={draw=none,legend to name=leg}
]
\addplot[smooth,mark=*,orange!50,mark size=0.75] table[y = number_of_nodes] from \dataTableNetworksize;
\addplot[smooth,mark=star,blue!50,mark size=0.75] table[y = number_of_edges] from \dataTableNetworksize;

   \addlegendentry{\# nodes}
   \addlegendentry{\# edges}

   \node [above left] (L) at (rel axis cs:0.7,0.6) {\ref{leg}};

\end{axis}
\end{tikzpicture}
\begin{tikzpicture}[scale=0.75]
\begin{axis}[
title={\# Added Edges},
xticklabels={,,},
]
\addplot[smooth,mark=*,green,mark size=0.75] table[y = batch] from \dataTableAddedEdges;
\addplot[smooth,mark=star,red,mark size=0.75] table[y = dynamic] from \dataTableAddedEdges;
\end{axis}
\end{tikzpicture}
\begin{tikzpicture}[scale=0.75]
\begin{axis}[
title={\# Centralities},
xticklabels={,,},
]
\addplot[smooth,mark=*,green,mark size=0.75] table[y = batch] from \dataTableNumberCentralities;
\addplot[smooth,mark=star,red,mark size=0.75] table[y = dynamic] from \dataTableNumberCentralities;
\end{axis}
\end{tikzpicture}
\\
\begin{tikzpicture}[scale=0.75]
\begin{axis}[
title={Speedup Ratio},
xticklabels={,,},
legend style={draw=none}
]
\addplot[smooth,mark=*,black!50,mark size=0.75] table[y = speedup] from \dataTableSpeedup;

\end{axis}
\end{tikzpicture}
\begin{tikzpicture}[scale=0.75]
\begin{axis}[
title={Elapsed Time},
xticklabels={,,},
]
\addplot[smooth,mark=*,green,mark size=0.75] table[y = batch] from \dataTableElapsedTime;
\addplot[smooth,mark=star,red,mark size=0.75] table[y = dynamic] from \dataTableElapsedTime;
\end{axis}
\end{tikzpicture}
\begin{tikzpicture}[scale=0.75]
\begin{axis}[
legend columns=-1,
legend entries={batch (incremental network), incremental (incremental network)},
legend to name=named,
title={Cumulative Time},
xticklabels={,,},
]
\addplot[smooth,mark=*,green,mark size=0.75] table[y = batch] from \dataTableCumulativeTime;
\addplot[smooth,mark=star,red,mark size=0.75] table[y = dynamic] from \dataTableCumulativeTime;
\end{axis}
\end{tikzpicture}
\\
\ref{named}
\end{center}

 \caption{Results for the DaysAll (incremental only, weighted)}
\label{fig:Results3}
\end{figure}
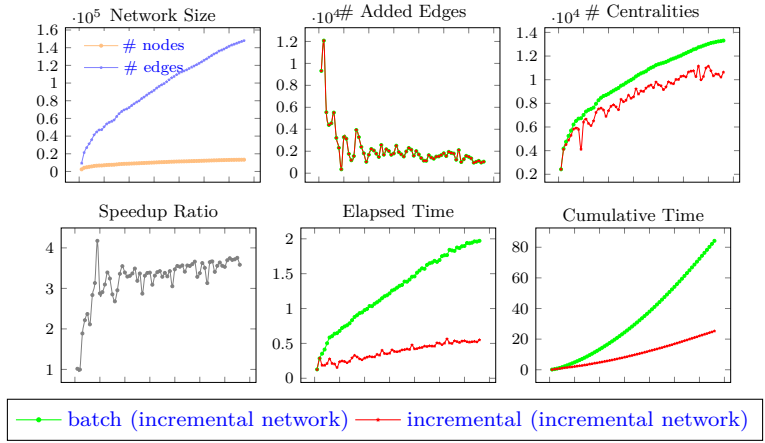

\begin{description}
    \item [\textbf{Unweighted Networks}:] The results presented in \figref{fig:Results1} are related to the High-energy physics theory citation network \cite{Leskovec2005} in an incremental unweighted network setting where no edges are removed from previous snapshots. \figref{fig:Results1} -- Network Size, shows the variation of the network over the 136 snapshots regarding the number of nodes and number of edges. \figref{fig:Results1} -- \# Added Edges, shows the number of added edges in each snapshot. Notice that the total number of edges in this dataset increases over the time. In the first snapshots a few edges are added to the network, but at the final snapshots, more than 6000 edges are added in each snapshot. \figref{fig:Results1} -- \# Centralities, shows that the number of calculated centralities in the incremental version is much lower than in the batch version. The batch version requires to compute centralities for all nodes in the snapshot. \figref{fig:Results1} -- Elapsed Time, shows the time required to perform the centralities measurements in each increment. This figure also shows that the incremental version is not only more efficient but also deals better with both increases in the size of network and increase in the number of added edges. This is confirmed by the total time required for processing the whole network: 27,806 seconds of the incremental version compared to the 121,345 seconds of the batch. It is clear that, as the number of added edges increases, the processing elapsed time of the batch version of the algorithm grows much faster than the incremental algorithm. Thus, regarding speed-up ratio, we achieve a speed-up of up to 6 times the processing time of the batch version with an incremental network (\figref{fig:Results1} -- Speedup Ratio). \\
    
    \item [\textbf{Weighted Networks}:] The results presented in \figref{fig:Results3} are related to the Reuters terror news network (DaysAll) \cite{Batagelj03densitybased} in an incremental weighted network setting, in this setup no edges are removed from previous snapshots. \figref{fig:Results3} -- Network Size, shows the variation of the network over the 66 snapshots regarding the number of nodes and number of edges, \figref{fig:Results3} -- \# Added Edges, shows the number of added edges in each snapshot, \figref{fig:Results3} -- \# Centralities, the number of calculated centralities in the incremental version (also much lower than in the batch version).The \figref{fig:Results3} -- Elapsed Time and (\figref{fig:Results3} -- Speedup Ratio) shown that were achieved a speed-up of up to 4 times the processing time of the batch version.

\end{description}

\subsection{Full Dynamic Network Results}
\label{TestScenario:DynamicNetworkSetup}

In this subsection, we provide results obtained with the full dynamic setup of the algorithm for unweighted and weighted networks.

\begin{figure}[!th]
  \centering

\pgfplotstableread[col sep=semicolon]{step;number_of_nodes;number_of_edges
1;2420;9318
2;4169;21396
3;4755;26943
4;5249;31337
5;5699;35883
6;6210;41402
7;6513;44616
8;6706;46930
9;6741;47283
10;7020;50563
11;7310;53705
12;7448;55468
13;7529;56646
14;7641;58202
15;7967;62137
16;8263;65419
17;8473;67811
18;8623;69620
19;8699;70658
20;8808;72342
21;8945;74551
22;9078;76621
23;9193;78405
24;9305;79884
25;9493;82450
26;9583;84133
27;9737;86310
28;9861;88330
29;9975;90006
30;10101;91803
31;10281;94307
32;10023;86909
33;9451;76550
34;9226;72523
35;9053;70106
36;9000;67856
37;8790;64480
38;8683;62858
39;8721;62553
40;8795;63908
41;8641;62001
42;8471;60002
43;8439;59371
44;8498;59830
45;8532;59701
46;8371;57087
47;8201;55316
48;8098;54454
49;8035;54288
50;8133;55064
51;8164;54939
52;8158;54676
53;8133;54452
54;8157;54456
55;8135;54204
56;8087;53739
57;8049;53072
58;8024;52167
59;8009;51727
60;7994;51541
61;7981;51094
62;7802;49551
63;7750;48664
64;7709;48069
65;7652;47518
66;7594;46592
67;7392;44296
68;7222;42153
69;7087;40561
70;6890;38552
71;6777;36844
72;6653;35471
73;6559;34328
74;6454;33196
75;6282;31559
76;6135;30132
77;6007;28811
78;5842;27300
79;5627;25770
80;5463;24127
81;5223;22313
82;5017;20754
83;4802;18808
84;4570;16962
85;4282;15174
86;4070;13947
87;3676;11846
88;3474;10830
89;3223;9558
90;2904;7978
91;2580;6488
92;2210;5138
93;1917;4177
94;1567;3144
95;1153;2020
}\dataTableNetworksize
\pgfplotstableread[col sep=semicolon]{step;batch;dynamic
1;9318;9318
2;12078;12078
3;5547;5547
4;4394;4394
5;4546;4546
6;5519;5519
7;3214;3214
8;2314;2314
9;353;353
10;3280;3280
11;3142;3142
12;1763;1763
13;1178;1178
14;1556;1556
15;3935;3935
16;3282;3282
17;2392;2392
18;1809;1809
19;1038;1038
20;1684;1684
21;2209;2209
22;2070;2070
23;1784;1784
24;1479;1479
25;2566;2566
26;1683;1683
27;2177;2177
28;2020;2020
29;1676;1676
30;1797;1797
31;2504;2504
32;-7398;1920
33;-10359;-7599
34;-4027;-10558
35;-2417;-3570
36;-2250;-2098
37;-3376;-2403
38;-1622;-3927
39;-305;-1205
40;1355;-606
41;-1907;1020
42;-1999;-2137
43;-631;-2010
44;459;-126
45;-129;249
46;-2614;-235
47;-1771;-2424
48;-862;-1752
49;-166;-749
50;776;5
51;-125;521
52;-263;262
53;-224;-363
54;4;-282
55;-252;-557
56;-465;622
57;-667;-1550
58;-905;-411
59;-440;-597
60;-186;-530
61;-447;-326
62;-1543;-836
63;-887;-1471
64;-595;-796
65;-551;-750
66;-926;-469
67;-2296;-1977
68;-2143;-2296
69;-1592;-2143
70;-2009;-1592
71;-1708;-2009
72;-1373;-1708
73;-1143;-1373
74;-1132;-1143
75;-1637;-1132
76;-1427;-1637
77;-1321;-1427
78;-1511;-1321
79;-1530;-1511
80;-1643;-1530
81;-1814;-1643
82;-1559;-1814
83;-1946;-1559
84;-1846;-1946
85;-1788;-1846
86;-1227;-1788
87;-2101;-1227
88;-1016;-2101
89;-1272;-1016
90;-1580;-1272
91;-1490;-1580
92;-1350;-1490
93;-961;-1350
94;-1033;-961
95;-1124;-1033
;;-1124
}\dataTableAddedEdges
\pgfplotstableread[col sep=semicolon]{step;batch;dynamic
1;2420;2420
2;4169;4115
3;4755;4517
4;5249;4892
5;5699;5293
6;6210;5825
7;6513;5901
8;6706;5803
9;6741;4114
10;7020;6415
11;7310;6656
12;7448;6303
13;7529;6124
14;7641;6512
15;7967;7229
16;8263;7516
17;8473;7590
18;8623;7436
19;8699;6901
20;8808;7378
21;8945;7748
22;9078;7872
23;9193;7709
24;9305;7462
25;9493;8332
26;9583;8134
27;9737;8259
28;9861;8685
29;9975;8430
30;10101;8547
31;10281;9222
32;10023;8785
33;9451;10017
34;9226;9828
35;9053;9086
36;9000;8969
37;8790;8824
38;8683;8739
39;8721;8419
40;8795;8101
41;8641;7676
42;8471;8269
43;8439;8157
44;8498;7852
45;8532;7762
46;8371;7830
47;8201;8090
48;8098;8004
49;8035;7749
50;8133;7606
51;8164;7404
52;8158;7651
53;8133;7729
54;8157;7758
55;8135;7439
56;8087;7611
57;8049;7617
58;8024;7419
59;8009;7509
60;7994;7557
61;7981;7439
62;7802;7310
63;7750;7401
64;7709;7110
65;7652;7042
66;7594;7031
67;7392;6665
68;7222;6668
69;7087;6441
70;6890;6058
71;6777;6195
72;6653;5758
73;6559;5665
74;6454;5326
75;6282;5353
76;6135;5448
77;6007;5220
78;5842;4988
79;5627;5012
80;5463;4827
81;5223;4740
82;5017;4645
83;4802;4414
84;4570;4351
85;4282;4096
86;4070;3897
87;3676;3513
88;3474;3575
89;3223;2871
90;2904;2728
91;2580;2763
92;2210;2462
93;1917;2132
94;1567;1731
95;1153;1584
;;1347
}\dataTableNumberCentralities
\pgfplotstableread[col sep=semicolon]{step;batch;dynamic
1;0.134;0.128
2;0.318;0.281
3;0.390;0.194
4;0.474;0.185
5;0.519;0.208
6;0.586;0.284
7;0.658;0.218
8;0.666;0.211
9;0.699;0.154
10;0.805;0.248
11;0.817;0.266
12;0.790;0.232
13;0.850;0.217
14;0.812;0.235
15;0.897;0.298
16;0.976;0.338
17;0.940;0.291
18;1.035;0.289
19;1.160;0.261
20;1.026;0.291
21;1.142;0.308
22;1.083;0.319
23;1.186;0.310
24;1.148;0.308
25;1.158;0.366
26;1.226;0.337
27;1.213;0.397
28;1.233;0.355
29;1.289;0.357
30;1.272;0.370
31;1.335;0.404
32;1.295;0.378
33;1.230;0.500
34;1.121;0.585
35;1.171;0.412
36;1.034;0.389
37;0.899;0.383
38;0.981;0.467
39;0.871;0.336
40;0.928;0.317
41;0.887;0.283
42;0.913;0.318
43;0.850;0.311
44;0.879;0.360
45;0.909;0.282
46;0.808;0.281
47;0.780;0.320
48;0.793;0.308
49;0.796;0.292
50;0.806;0.284
51;0.792;0.327
52;0.803;0.292
53;0.821;0.295
54;0.756;0.302
55;0.755;0.288
56;0.782;0.281
57;0.779;0.278
58;0.787;0.260
59;0.729;0.326
60;0.773;0.278
61;0.733;0.268
62;0.757;0.264
63;0.723;0.278
64;0.743;0.259
65;0.684;0.243
66;0.700;0.242
67;0.624;0.219
68;0.595;0.274
69;0.610;0.211
70;0.552;0.193
71;0.548;0.193
72;0.514;0.172
73;0.498;0.177
74;0.476;0.163
75;0.465;0.161
76;0.433;0.160
77;0.407;0.149
78;0.440;0.138
79;0.358;0.136
80;0.348;0.138
81;0.375;0.133
82;0.335;0.129
83;0.269;0.112
84;0.263;0.116
85;0.213;0.111
86;0.230;0.101
87;0.174;0.086
88;0.158;0.127
89;0.133;0.065
90;0.119;0.064
91;0.097;0.067
92;0.075;0.058
93;0.062;0.050
94;0.048;0.038
95;0.033;0.037
;;0.032
}\dataTableElapsedTime
\pgfplotstableread[col sep=semicolon]{step;batch;dynamic
1;0.134;0.128
2;0.452;0.409
3;0.842;0.603
4;1.316;0.788
5;1.835;0.996
6;2.421;1.28
7;3.079;1.498
8;3.745;1.709
9;4.444;1.863
10;5.249;2.111
11;6.066;2.377
12;6.856;2.609
13;7.706;2.826
14;8.518;3.061
15;9.415;3.359
16;10.391;3.697
17;11.331;3.988
18;12.366;4.277
19;13.526;4.538
20;14.552;4.829
21;15.694;5.137
22;16.777;5.456
23;17.963;5.766
24;19.111;6.074
25;20.269;6.44
26;21.495;6.777
27;22.708;7.174
28;23.941;7.529
29;25.23;7.886
30;26.502;8.256
31;27.837;8.66
32;29.132;9.038
33;30.362;9.538
34;31.483;10.123
35;32.654;10.535
36;33.688;10.924
37;34.587;11.307
38;35.568;11.774
39;36.439;12.11
40;37.367;12.427
41;38.254;12.71
42;39.167;13.028
43;40.017;13.339
44;40.896;13.699
45;41.805;13.981
46;42.613;14.262
47;43.393;14.582
48;44.186;14.89
49;44.982;15.182
50;45.788;15.466
51;46.58;15.793
52;47.383;16.085
53;48.204;16.38
54;48.96;16.682
55;49.715;16.97
56;50.497;17.251
57;51.276;17.529
58;52.063;17.789
59;52.792;18.115
60;53.565;18.393
61;54.298;18.661
62;55.055;18.925
63;55.778;19.203
64;56.521;19.462
65;57.205;19.705
66;57.905;19.947
67;58.529;20.166
68;59.124;20.44
69;59.734;20.651
70;60.286;20.844
71;60.834;21.037
72;61.348;21.209
73;61.846;21.386
74;62.322;21.549
75;62.787;21.71
76;63.22;21.87
77;63.627;22.019
78;64.067;22.157
79;64.425;22.293
80;64.773;22.431
81;65.148;22.564
82;65.483;22.693
83;65.752;22.805
84;66.015;22.921
85;66.228;23.032
86;66.458;23.133
87;66.632;23.219
88;66.79;23.346
89;66.923;23.411
90;67.042;23.475
91;67.139;23.542
92;67.214;23.6
93;67.276;23.65
94;67.324;23.688
95;67.357;23.725
;;23.757
}\dataTableCumulativeTime
\pgfplotstableread[col sep=semicolon]{step;speedup
1;1.04688
2;1.13167
3;2.01031
4;2.56216
5;2.49519
6;2.06338
7;3.01835
8;3.1564
9;4.53896
10;3.24597
11;3.07143
12;3.40517
13;3.91705
14;3.45532
15;3.01007
16;2.88757
17;3.23024
18;3.58131
19;4.44444
20;3.52577
21;3.70779
22;3.39498
23;3.82581
24;3.72727
25;3.16393
26;3.63798
27;3.05542
28;3.47324
29;3.61064
30;3.43784
31;3.30446
32;3.42593
33;2.46
34;1.91624
35;2.84223
36;2.6581
37;2.34726
38;2.10064
39;2.59226
40;2.92744
41;3.13428
42;2.87107
43;2.73312
44;2.44167
45;3.2234
46;2.87544
47;2.4375
48;2.57468
49;2.72603
50;2.83803
51;2.42202
52;2.75
53;2.78305
54;2.50331
55;2.62153
56;2.78292
57;2.80216
58;3.02692
59;2.2362
60;2.78058
61;2.73507
62;2.86742
63;2.60072
64;2.86873
65;2.81481
66;2.89256
67;2.84932
68;2.17153
69;2.891
70;2.8601
71;2.83938
72;2.98837
73;2.81356
74;2.92025
75;2.8882
76;2.70625
77;2.73154
78;3.18841
79;2.63235
80;2.52174
81;2.81955
82;2.5969
83;2.40179
84;2.26724
85;1.91892
86;2.27723
87;2.02326
88;1.24409
89;2.04615
90;1.85937
91;1.44776
92;1.2931
93;1.24
94;1.26316
95;0.891892
}\dataTableSpeedup

\pgfplotsset{footnotesize}
\begin{center}
%
%
\begin{tikzpicture}[scale=0.75]
\begin{axis}[
title={Network Size},
xticklabels={,,},
legend style={draw=none,legend to name=leg}
]
\addplot[smooth,mark=*,orange!50,mark size=0.75] table[y = number_of_nodes] from \dataTableNetworksize;
\addplot[smooth,mark=star,blue!50,mark size=0.75] table[y = number_of_edges] from \dataTableNetworksize;

   \addlegendentry{\# nodes}
   \addlegendentry{\# edges}

   \node [above left] (L) at (rel axis cs:0.7,0.6) {\ref{leg}};

\end{axis}
\end{tikzpicture}
\begin{tikzpicture}[scale=0.75]
\begin{axis}[
title={\# Added Edges},
xticklabels={,,},
]
\addplot[smooth,mark=*,green,mark size=0.75] table[y = batch] from \dataTableAddedEdges;
\addplot[smooth,mark=star,red,mark size=0.75] table[y = dynamic] from \dataTableAddedEdges;
\end{axis}
\end{tikzpicture}
\begin{tikzpicture}[scale=0.75]
\begin{axis}[
title={\# Centralities},
xticklabels={,,},
]
\addplot[smooth,mark=*,green,mark size=0.75] table[y = batch] from \dataTableNumberCentralities;
\addplot[smooth,mark=star,red,mark size=0.75] table[y = dynamic] from \dataTableNumberCentralities;
\end{axis}
\end{tikzpicture}
\\
\begin{tikzpicture}[scale=0.75]
\begin{axis}[
title={Speedup Ratio},
xticklabels={,,},
legend style={draw=none}
]
\addplot[smooth,mark=*,black!50,mark size=0.75] table[y = speedup] from \dataTableSpeedup;

\end{axis}
\end{tikzpicture}
\begin{tikzpicture}[scale=0.75]
\begin{axis}[
title={Elapsed Time},
xticklabels={,,},
]
\addplot[smooth,mark=*,green,mark size=0.75] table[y = batch] from \dataTableElapsedTime;
\addplot[smooth,mark=star,red,mark size=0.75] table[y = dynamic] from \dataTableElapsedTime;
\end{axis}
\end{tikzpicture}
\begin{tikzpicture}[scale=0.75]
\begin{axis}[
legend columns=-1,
legend entries={batch (dynamic network), incremental (dynamic network)},
legend to name=named,
title={Cumulative Time},
xticklabels={,,},
]
\addplot[smooth,mark=*,green,mark size=0.75] table[y = batch] from \dataTableCumulativeTime;
\addplot[smooth,mark=star,red,mark size=0.75] table[y = dynamic] from \dataTableCumulativeTime;
\end{axis}
\end{tikzpicture}
\\
\ref{named}
\end{center}

 \caption{Results for the DaysAll network (full dynamic, weighted)}
\label{fig:Results5}
\end{figure}
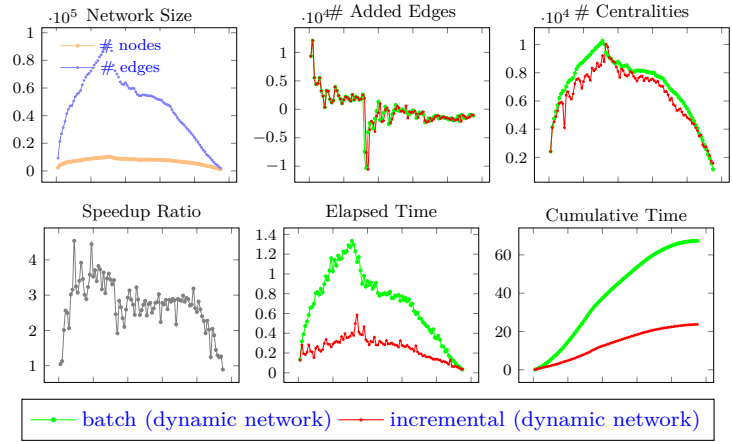

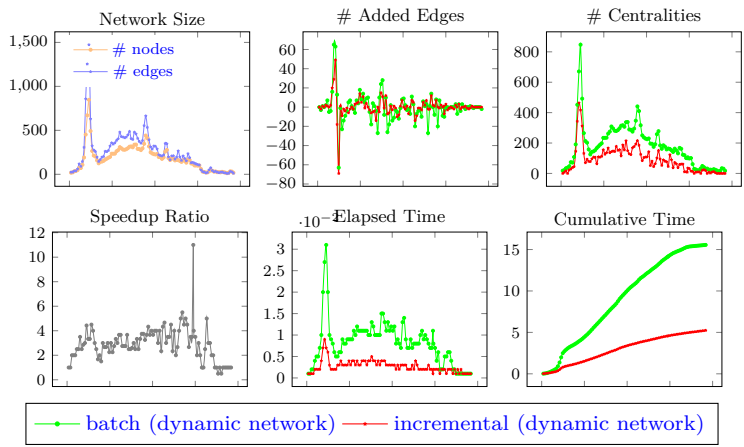
\begin{figure}[!th]
  \centering

\pgfplotstableread[col sep=semicolon]{step;number_of_nodes;number_of_edges
19;18;24
38;22;25
57;34;37
76;38;39
95;47;61
114;74;119
133;71;91
152;108;147
171;191;306
190;452;858
209;671;1254
228;847;1474
247;492;630
266;265;323
285;213;275
304;202;259
323;158;194
342;125;146
361;142;168
380;145;194
399;153;203
418;167;261
437;178;258
456;189;271
475;198;321
494;234;358
513;234;343
532;235;358
551;262;386
570;275;404
589;309;479
608;296;428
627;281;424
646;279;415
665;285;423
684;297;448
703;314;488
722;317;406
741;307;397
760;337;479
779;338;459
798;289;409
817;287;374
836;275;354
855;295;385
874;347;554
893;441;664
912;410;534
931;317;394
950;250;312
969;237;286
988;229;322
1007;225;318
1026;230;288
1045;196;223
1064;195;213
1083;175;209
1102;192;219
1121;258;351
1140;278;347
1159;184;198
1178;162;166
1197;185;209
1216;176;211
1235;158;173
1254;164;195
1273;161;266
1292;152;197
1311;141;158
1330;163;209
1349;151;180
1368;132;153
1387;168;235
1406;162;190
1425;123;133
1444;120;135
1463;101;126
1477;100;118
1482;106;119
1501;98;111
1520;101;105
1539;61;50
1558;30;31
1577;26;25
1596;20;16
1615;21;21
1634;33;51
1653;49;75
1672;41;43
1691;30;30
1710;32;33
1729;24;25
1748;25;19
1767;22;23
1786;28;26
1805;24;18
1824;19;17
1843;13;10
1862;17;11
1881;34;32
1900;31;28
1919;18;12
}\dataTableNetworksize
\pgfplotstableread[col sep=semicolon]{step;batch;dynamic
19;0;0
38;-3;-2
57;1;1
76;0;0
95;2;2
114;7;1
133;-5;0
152;-4;2
171;16;20
190;65;29
209;63;49
228;13;-18
247;-63;-69
266;-1;1
285;-23;-9
304;-14;-2
323;-9;-4
342;-4;-1
361;5;-6
380;7;-2
399;1;0
418;-2;-2
437;-6;3
456;7;4
475;4;5
494;18;14
513;8;5
532;14;11
551;5;-1
570;9;4
589;10;0
608;-5;-3
627;-18;-6
646;4;-5
665;1;-4
684;-4;-14
703;-27;0
722;10;-5
741;24;15
760;28;11
779;-8;-1
798;10;12
817;-26;-7
836;-18;-6
855;2;1
874;-14;-3
893;-11;1
912;-3;-1
931;-5;-10
950;-12;-8
969;-1;1
988;8;7
1007;9;0
1026;2;-2
1045;-3;1
1064;-2;-7
1083;-7;0
1102;11;7
1121;3;5
1140;-20;-14
1159;-7;-9
1178;-4;-1
1197;-3;-1
1216;-10;-3
1235;6;5
1254;1;2
1273;3;7
1292;-27;-2
1311;1;-2
1330;14;12
1349;0;2
1368;-8;1
1387;8;7
1406;-20;0
1425;7;-1
1444;0;2
1463;-9;0
1477;0;1
1482;-10;-7
1501;-1;3
1520;0;-3
1539;-1;-1
1558;0;0
1577;0;-1
1596;0;-2
1615;3;2
1634;-7;0
1653;3;3
1672;-4;-2
1691;3;0
1710;0;0
1729;-2;1
1748;0;-1
1767;0;-2
1786;1;-1
1805;0;1
1824;-1;0
1843;0;0
1862;0;0
1881;0;0
1900;0;0
1919;-2;0
}\dataTableAddedEdges
\pgfplotstableread[col sep=semicolon]{step;batch;dynamic
19;18;0
38;22;7
57;34;17
76;38;0
95;47;22
114;74;29
133;71;39
152;108;45
171;191;146
190;452;343
209;671;464
228;847;417
247;492;312
266;265;132
285;213;70
304;202;93
323;158;79
342;125;78
361;142;96
380;145;107
399;153;106
418;167;63
437;178;109
456;189;107
475;198;113
494;234;156
513;234;126
532;235;136
551;262;167
570;275;136
589;309;143
608;296;147
627;281;144
646;279;178
665;285;152
684;297;156
703;314;115
722;317;187
741;307;154
760;337;215
779;338;150
798;289;129
817;287;124
836;275;169
855;295;177
874;347;193
893;441;216
912;410;185
931;317;126
950;250;84
969;237;105
988;229;107
1007;225;109
1026;230;143
1045;196;52
1064;195;80
1083;175;45
1102;192;64
1121;258;101
1140;278;151
1159;184;94
1178;162;53
1197;185;89
1216;176;80
1235;158;77
1254;164;46
1273;161;123
1292;152;90
1311;141;37
1330;163;72
1349;151;63
1368;132;38
1387;168;59
1406;162;61
1425;123;61
1444;120;10
1463;101;12
1477;100;8
1482;106;35
1501;98;32
1520;101;15
1539;61;13
1558;30;0
1577;26;9
1596;20;9
1615;21;11
1634;33;0
1653;49;21
1672;41;25
1691;30;19
1710;32;0
1729;24;8
1748;25;2
1767;22;9
1786;28;12
1805;24;5
1824;19;0
1843;13;0
1862;17;0
1881;34;0
1900;31;0
1919;18;0
}\dataTableNumberCentralities
\pgfplotstableread[col sep=semicolon]{step;batch;dynamic
19;0.001;0.001
38;0.001;0.001
57;0.002;0.001
76;0.002;0.001
95;0.004;0.002
114;0.005;0.002
133;0.005;0.002
152;0.007;0.002
171;0.010;0.004
190;0.020;0.007
209;0.027;0.009
228;0.031;0.007
247;0.020;0.006
266;0.010;0.003
285;0.009;0.002
304;0.008;0.002
323;0.006;0.002
342;0.005;0.002
361;0.005;0.003
380;0.006;0.003
399;0.006;0.004
418;0.009;0.003
437;0.008;0.003
456;0.008;0.003
475;0.009;0.003
494;0.009;0.004
513;0.009;0.003
532;0.009;0.004
551;0.011;0.004
570;0.010;0.003
589;0.012;0.004
608;0.011;0.003
627;0.011;0.003
646;0.011;0.004
665;0.011;0.004
684;0.011;0.004
703;0.011;0.003
722;0.010;0.004
741;0.010;0.004
760;0.013;0.005
779;0.012;0.004
798;0.010;0.004
817;0.010;0.003
836;0.010;0.004
855;0.010;0.003
874;0.015;0.004
893;0.015;0.004
912;0.013;0.004
931;0.011;0.003
950;0.013;0.003
969;0.011;0.003
988;0.012;0.003
1007;0.011;0.003
1026;0.012;0.003
1045;0.008;0.002
1064;0.009;0.003
1083;0.008;0.002
1102;0.007;0.003
1121;0.013;0.003
1140;0.014;0.003
1159;0.009;0.003
1178;0.007;0.002
1197;0.007;0.002
1216;0.009;0.002
1235;0.007;0.003
1254;0.008;0.002
1273;0.008;0.004
1292;0.008;0.003
1311;0.008;0.002
1330;0.010;0.002
1349;0.011;0.002
1368;0.009;0.002
1387;0.010;0.002
1406;0.009;0.002
1425;0.008;0.003
1444;0.007;0.002
1463;0.007;0.002
1477;0.011;0.001
1482;0.008;0.002
1501;0.007;0.002
1520;0.004;0.002
1539;0.006;0.002
1558;0.002;0.001
1577;0.002;0.002
1596;0.002;0.002
1615;0.005;0.002
1634;0.005;0.001
1653;0.006;0.002
1672;0.006;0.002
1691;0.004;0.002
1710;0.002;0.001
1729;0.002;0.002
1748;0.001;0.001
1767;0.001;0.002
1786;0.001;0.001
1805;0.001;0.002
1824;0.001;0.001
1843;0.001;0.001
1862;0.001;0.001
1881;0.001;0.001
1900;0.001;0.001
1919;0.001;0.001
}\dataTableElapsedTime
\pgfplotstableread[col sep=semicolon]{step;batch;dynamic
19;0.02;0.023
38;0.045;0.044
57;0.08;0.065
76;0.139;0.089
95;0.211;0.118
114;0.3;0.156
133;0.396;0.193
152;0.497;0.234
171;0.636;0.283
190;0.912;0.373
209;1.366;0.503
228;1.988;0.668
247;2.493;0.797
266;2.729;0.869
285;2.912;0.924
304;3.073;0.975
323;3.211;1.023
342;3.324;1.065
361;3.423;1.11
380;3.537;1.162
399;3.653;1.211
418;3.79;1.27
437;3.932;1.325
456;4.067;1.387
475;4.219;1.442
494;4.391;1.501
513;4.568;1.571
532;4.744;1.641
551;4.937;1.706
570;5.145;1.773
589;5.355;1.838
608;5.575;1.907
627;5.786;1.978
646;5.993;2.045
665;6.199;2.116
684;6.413;2.194
703;6.625;2.268
722;6.841;2.341
741;7.04;2.408
760;7.248;2.481
779;7.476;2.561
798;7.688;2.644
817;7.89;2.714
836;8.086;2.775
855;8.277;2.839
874;8.497;2.913
893;8.767;3.005
912;9.042;3.092
931;9.266;3.172
950;9.478;3.235
969;9.682;3.294
988;9.895;3.353
1007;10.102;3.415
1026;10.305;3.475
1045;10.479;3.53
1064;10.637;3.581
1083;10.791;3.627
1102;10.95;3.671
1121;11.142;3.728
1140;11.373;3.787
1159;11.559;3.841
1178;11.698;3.883
1197;11.84;3.925
1216;12.008;3.969
1235;12.162;4.013
1254;12.298;4.054
1273;12.453;4.104
1292;12.613;4.153
1311;12.752;4.196
1330;12.898;4.241
1349;13.056;4.287
1368;13.235;4.336
1387;13.411;4.383
1406;13.597;4.432
1425;13.752;4.474
1444;13.894;4.511
1463;14.038;4.548
1477;14.144;4.575
1482;14.184;4.585
1501;14.318;4.623
1520;14.445;4.662
1539;14.568;4.696
1558;14.652;4.727
1577;14.74;4.756
1596;14.823;4.785
1615;14.897;4.818
1634;14.983;4.848
1653;15.078;4.877
1672;15.176;4.911
1691;15.268;4.939
1710;15.318;4.965
1729;15.364;4.992
1748;15.385;5.015
1767;15.404;5.043
1786;15.423;5.069
1805;15.442;5.092
1824;15.461;5.117
1843;15.48;5.142
1862;15.499;5.165
1881;15.518;5.19
1900;15.537;5.217
1919;15.556;5.243
}\dataTableCumulativeTime
\pgfplotstableread[col sep=semicolon]{step;speedup
19;1
38;1
57;2
76;2
95;2
114;2.5
133;2.5
152;3.5
171;2.5
190;2.85714
209;3
228;4.42857
247;3.33333
266;3.33333
285;4.5
304;4
323;3
342;2.5
361;1.66667
380;2
399;1.5
418;3
437;2.66667
456;2.66667
475;3
494;2.25
513;3
532;2.25
551;2.75
570;3.33333
589;3
608;3.66667
627;3.66667
646;2.75
665;2.75
684;2.75
703;3.66667
722;2.5
741;2.5
760;2.6
779;3
798;2.5
817;3.33333
836;2.5
855;3.33333
874;3.75
893;3.75
912;3.25
931;3.66667
950;4.33333
969;3.66667
988;4
1007;3.66667
1026;4
1045;4
1064;3
1083;4
1102;2.33333
1121;4.33333
1140;4.66667
1159;3
1178;3.5
1197;3.5
1216;4.5
1235;2.33333
1254;4
1273;2
1292;2.66667
1311;4
1330;5
1349;5.5
1368;4.5
1387;5
1406;4.5
1425;2.66667
1444;3.5
1463;3.5
1477;11
1482;4
1501;3.5
1520;2
1539;3
1558;2
1577;1
1596;1
1615;2.5
1634;5
1653;3
1672;3
1691;2
1710;2
1729;1
1748;1
1767;0.5
1786;1
1805;0.5
1824;1
1843;1
1862;1
1881;1
1900;1
1919;1
}\dataTableSpeedup

\pgfplotsset{footnotesize}
\begin{center}
%
%
\begin{tikzpicture}[scale=0.75]
\begin{axis}[
title={Network Size},
xticklabels={,,},
legend style={draw=none,legend to name=leg1}
]
\addplot[smooth,mark=*,orange!50,mark size=0.75] table[y = number_of_nodes] from \dataTableNetworksize;
\addplot[smooth,mark=star,blue!50,mark size=0.75] table[y = number_of_edges] from \dataTableNetworksize;

   \addlegendentry{\# nodes}
   \addlegendentry{\# edges}

   \node [above left] (L) at (rel axis cs:0.7,0.6) {\ref{leg1}};

\end{axis}
\end{tikzpicture}
\begin{tikzpicture}[scale=0.75]
\begin{axis}[
title={\# Added Edges},
xticklabels={,,},
]
\addplot[smooth,mark=*,green,mark size=0.75] table[y = batch] from \dataTableAddedEdges;
\addplot[smooth,mark=star,red,mark size=0.75] table[y = dynamic] from \dataTableAddedEdges;
\end{axis}
\end{tikzpicture}
\begin{tikzpicture}[scale=0.75]
\begin{axis}[
title={\# Centralities},
xticklabels={,,},
]
\addplot[smooth,mark=*,green,mark size=0.75] table[y = batch] from \dataTableNumberCentralities;
\addplot[smooth,mark=star,red,mark size=0.75] table[y = dynamic] from \dataTableNumberCentralities;
\end{axis}
\end{tikzpicture}
\\
\begin{tikzpicture}[scale=0.75]
\begin{axis}[
title={Speedup Ratio},
xticklabels={,,},
legend style={draw=none}
]
\addplot[smooth,mark=*,black!50,mark size=0.75] table[y = speedup] from \dataTableSpeedup;

\end{axis}
\end{tikzpicture}
\begin{tikzpicture}[scale=0.75]
\begin{axis}[
title={Elapsed Time},
xticklabels={,,},
]
\addplot[smooth,mark=*,green,mark size=0.75] table[y = batch] from \dataTableElapsedTime;
\addplot[smooth,mark=star,red,mark size=0.75] table[y = dynamic] from \dataTableElapsedTime;
\end{axis}
\end{tikzpicture}
\begin{tikzpicture}[scale=0.75]
\begin{axis}[
legend columns=-1,
legend entries={batch (dynamic network), incremental (dynamic network)},
legend to name=named,
title={Cumulative Time},
xticklabels={,,},
]
\addplot[smooth,mark=*,green,mark size=0.75] table[y = batch] from \dataTableCumulativeTime;
\addplot[smooth,mark=star,red,mark size=0.75] table[y = dynamic] from \dataTableCumulativeTime;
\end{axis}
\end{tikzpicture}
\\
\ref{named}
\end{center}

 \caption{Results for the Bitcoinalpha (full dynamic, 30 day sliding window, weighted)}
\label{fig:Results6}
\end{figure}

\begin{figure}[!th]
  \centering

\pgfplotstableread[col sep=semicolon]{step;number_of_nodes;number_of_edges
1;41;62
2;60;99
3;97;201
4;142;303
5;278;624
6;678;1681
7;1197;3014
8;1289;3310
9;1362;3583
10;1407;3740
11;1456;3901
12;1518;4069
13;1590;4294
14;1655;4504
15;1725;4759
16;1816;4974
17;1890;5200
18;1948;5263
19;1770;4595
20;1304;3662
21;1262;3734
22;1254;3899
23;1322;4121
24;1383;4349
25;1427;4584
26;1464;4696
27;1481;4751
28;1545;4985
29;1635;5166
30;1672;5180
31;1653;5054
32;1613;4939
33;1600;4797
34;1566;4598
35;1516;4359
36;1496;4286
37;1449;4133
38;1386;3888
39;1347;3725
40;1320;3598
41;1243;3315
42;1110;2973
43;1049;2815
44;1008;2708
45;998;2635
46;953;2484
47;918;2422
48;880;2315
49;804;2108
50;723;1894
51;667;1757
52;609;1587
53;552;1495
54;518;1318
55;479;1201
56;435;1051
57;387;926
58;326;718
59;285;608
60;250;491
61;224;416
62;162;325
63;142;296
64;135;275
65;127;253
66;118;206
67;100;165
68;91;140
69;82;117
70;70;94
71;59;72
72;50;55
73;46;47
}\dataTableNetworksize
\pgfplotstableread[col sep=semicolon]{step;batch;dynamic
1;98;62
2;67;41
3;188;108
4;175;108
5;548;333
6;1896;1086
7;2424;1375
8;553;323
9;490;285
10;282;169
11;256;167
12;296;181
13;387;261
14;331;226
15;424;284
16;365;300
17;392;245
18;71;296
19;-1227;85
20;-1738;-654
21;89;-956
22;185;153
23;354;133
24;388;255
25;384;266
26;160;207
27;81;40
28;330;299
29;277;208
30;11;27
31;-192;-82
32;-169;-88
33;-250;-135
34;-348;-240
35;-334;-201
36;-148;-155
37;-295;-158
38;-427;-253
39;-294;-219
40;-280;-242
41;-488;-55
42;-608;-387
43;-307;-326
44;-219;-209
45;-184;-30
46;-260;-186
47;-120;-148
48;-191;-77
49;-339;-110
50;-314;-290
51;-230;-237
52;-270;-148
53;-129;-163
54;-289;-102
55;-191;-216
56;-242;-142
57;-184;-178
58;-291;-136
59;-156;-224
60;-165;-121
61;-112;-103
62;-132;-99
63;-35;-108
64;-23;-31
65;-24;-23
66;-60;-23
67;-44;-51
68;-31;-44
69;-29;-30
70;-24;-25
71;-25;-23
72;-20;-23
73;-9;-17
;;-9
}\dataTableAddedEdges
\pgfplotstableread[col sep=semicolon]{step;batch;dynamic
1;41;41
2;60;56
3;97;90
4;142;131
5;278;274
6;678;672
7;1197;1149
8;1289;893
9;1362;905
10;1407;734
11;1456;849
12;1518;760
13;1590;882
14;1655;1087
15;1725;1130
16;1816;1138
17;1890;1324
18;1948;1425
19;1770;1533
20;1304;1824
21;1262;1784
22;1254;1321
23;1322;1290
24;1383;1283
25;1427;1288
26;1464;1382
27;1481;1395
28;1545;1505
29;1635;1627
30;1672;1601
31;1653;1481
32;1613;1522
33;1600;1520
34;1566;1463
35;1516;1454
36;1496;1449
37;1449;1442
38;1386;1386
39;1347;1317
40;1320;1250
41;1243;1191
42;1110;1202
43;1049;1174
44;1008;1046
45;998;955
46;953;934
47;918;889
48;880;820
49;804;780
50;723;762
51;667;697
52;609;597
53;552;565
54;518;520
55;479;488
56;435;444
57;387;431
58;326;377
59;285;351
60;250;300
61;224;274
62;162;243
63;142;207
64;135;141
65;127;128
66;118;108
67;100;114
68;91;105
69;82;80
70;70;74
71;59;61
72;50;60
73;46;47
;;32
}\dataTableNumberCentralities
\pgfplotstableread[col sep=semicolon]{step;batch;dynamic
1;0.002;0.002
2;0.002;0.003
3;0.004;0.005
4;0.008;0.005
5;0.012;0.010
6;0.033;0.030
7;0.063;0.040
8;0.064;0.019
9;0.068;0.021
10;0.076;0.016
11;0.078;0.016
12;0.082;0.016
13;0.090;0.018
14;0.091;0.021
15;0.097;0.024
16;0.098;0.025
17;0.107;0.029
18;0.104;0.030
19;0.088;0.036
20;0.071;0.051
21;0.074;0.054
22;0.069;0.045
23;0.075;0.030
24;0.079;0.028
25;0.082;0.031
26;0.086;0.030
27;0.089;0.030
28;0.092;0.035
29;0.097;0.039
30;0.126;0.035
31;0.095;0.033
32;0.099;0.036
33;0.092;0.036
34;0.090;0.033
35;0.086;0.031
36;0.090;0.033
37;0.075;0.033
38;0.076;0.029
39;0.075;0.029
40;0.078;0.027
41;0.062;0.025
42;0.068;0.029
43;0.057;0.027
44;0.051;0.023
45;0.046;0.022
46;0.047;0.021
47;0.046;0.020
48;0.042;0.016
49;0.035;0.016
50;0.032;0.016
51;0.033;0.014
52;0.026;0.012
53;0.025;0.012
54;0.023;0.010
55;0.021;0.011
56;0.017;0.010
57;0.016;0.010
58;0.012;0.008
59;0.010;0.009
60;0.008;0.007
61;0.007;0.006
62;0.005;0.006
63;0.005;0.005
64;0.005;0.003
65;0.004;0.003
66;0.003;0.003
67;0.003;0.003
68;0.002;0.003
69;0.002;0.003
70;0.002;0.002
71;0.002;0.002
72;0.001;0.002
73;0.001;0.002
;;0.002
}\dataTableElapsedTime
\pgfplotstableread[col sep=semicolon]{step;batch;dynamic
1;0.002;0.002
2;0.004;0.005
3;0.008;0.01
4;0.016;0.015
5;0.028;0.025
6;0.061;0.055
7;0.124;0.095
8;0.188;0.114
9;0.256;0.135
10;0.332;0.151
11;0.41;0.167
12;0.492;0.183
13;0.582;0.201
14;0.673;0.222
15;0.77;0.246
16;0.868;0.271
17;0.975;0.3
18;1.079;0.33
19;1.167;0.366
20;1.238;0.417
21;1.312;0.471
22;1.381;0.516
23;1.456;0.546
24;1.535;0.574
25;1.617;0.605
26;1.703;0.635
27;1.792;0.665
28;1.884;0.7
29;1.981;0.739
30;2.107;0.774
31;2.202;0.807
32;2.301;0.843
33;2.393;0.879
34;2.483;0.912
35;2.569;0.943
36;2.659;0.976
37;2.734;1.009
38;2.81;1.038
39;2.885;1.067
40;2.963;1.094
41;3.025;1.119
42;3.093;1.148
43;3.15;1.175
44;3.201;1.198
45;3.247;1.22
46;3.294;1.241
47;3.34;1.261
48;3.382;1.277
49;3.417;1.293
50;3.449;1.309
51;3.482;1.323
52;3.508;1.335
53;3.533;1.347
54;3.556;1.357
55;3.577;1.368
56;3.594;1.378
57;3.61;1.388
58;3.622;1.396
59;3.632;1.405
60;3.64;1.412
61;3.647;1.418
62;3.652;1.424
63;3.657;1.429
64;3.662;1.432
65;3.666;1.435
66;3.669;1.438
67;3.672;1.441
68;3.674;1.444
69;3.676;1.447
70;3.678;1.449
71;3.68;1.451
72;3.681;1.453
73;3.682;1.455
;;1.457
}\dataTableCumulativeTime
\pgfplotstableread[col sep=semicolon]{step;speedup
1;1
2;0.666667
3;0.8
4;1.6
5;1.2
6;1.1
7;1.575
8;3.36842
9;3.2381
10;4.75
11;4.875
12;5.125
13;5
14;4.33333
15;4.04167
16;3.92
17;3.68966
18;3.46667
19;2.44444
20;1.39216
21;1.37037
22;1.53333
23;2.5
24;2.82143
25;2.64516
26;2.86667
27;2.96667
28;2.62857
29;2.48718
30;3.6
31;2.87879
32;2.75
33;2.55556
34;2.72727
35;2.77419
36;2.72727
37;2.27273
38;2.62069
39;2.58621
40;2.88889
41;2.48
42;2.34483
43;2.11111
44;2.21739
45;2.09091
46;2.2381
47;2.3
48;2.625
49;2.1875
50;2
51;2.35714
52;2.16667
53;2.08333
54;2.3
55;1.90909
56;1.7
57;1.6
58;1.5
59;1.11111
60;1.14286
61;1.16667
62;0.833333
63;1
64;1.66667
65;1.33333
66;1
67;1
68;0.666667
69;0.666667
70;1
71;1
72;0.5
73;0.5
}\dataTableSpeedup

\pgfplotsset{footnotesize}
\begin{center}
%
%
\begin{tikzpicture}[scale=0.75]
\begin{axis}[
title={Network Size},
xticklabels={,,},
legend style={draw=none,legend to name=leg}
]
\addplot[smooth,mark=*,orange!50,mark size=0.75] table[y = number_of_nodes] from \dataTableNetworksize;
\addplot[smooth,mark=star,blue!50,mark size=0.75] table[y = number_of_edges] from \dataTableNetworksize;

   \addlegendentry{\# nodes}
   \addlegendentry{\# edges}

   \node [above left] (L) at (rel axis cs:0.7,0.6) {\ref{leg}};

\end{axis}
\end{tikzpicture}
\begin{tikzpicture}[scale=0.75]
\begin{axis}[
title={\# Added Edges},
xticklabels={,,},
]
\addplot[smooth,mark=*,green,mark size=0.75] table[y = batch] from \dataTableAddedEdges;
\addplot[smooth,mark=star,red,mark size=0.75] table[y = dynamic] from \dataTableAddedEdges;
\end{axis}
\end{tikzpicture}
\begin{tikzpicture}[scale=0.75]
\begin{axis}[
title={\# Centralities},
xticklabels={,,},
]
\addplot[smooth,mark=*,green,mark size=0.75] table[y = batch] from \dataTableNumberCentralities;
\addplot[smooth,mark=star,red,mark size=0.75] table[y = dynamic] from \dataTableNumberCentralities;
\end{axis}
\end{tikzpicture}
\\
\begin{tikzpicture}[scale=0.75]
\begin{axis}[
title={Speedup Ratio},
xticklabels={,,},
legend style={draw=none}
]
\addplot[smooth,mark=*,black!50,mark size=0.75] table[y = speedup] from \dataTableSpeedup;

\end{axis}
\end{tikzpicture}
\begin{tikzpicture}[scale=0.75]
\begin{axis}[
title={Elapsed Time},
xticklabels={,,},
]
\addplot[smooth,mark=*,green,mark size=0.75] table[y = batch] from \dataTableElapsedTime;
\addplot[smooth,mark=star,red,mark size=0.75] table[y = dynamic] from \dataTableElapsedTime;
\end{axis}
\end{tikzpicture}
\begin{tikzpicture}[scale=0.75]
\begin{axis}[
legend columns=-1,
legend entries={batch (dynamic network), incremental (dynamic network)},
legend to name=named,
title={Cumulative Time},
xticklabels={,,},
]
\addplot[smooth,mark=*,green,mark size=0.75] table[y = batch] from \dataTableCumulativeTime;
\addplot[smooth,mark=star,red,mark size=0.75] table[y = dynamic] from \dataTableCumulativeTime;
\end{axis}
\end{tikzpicture}
\\
\ref{named}
\end{center}

 \caption{Results for the Bitcoinalpha Month (full dynamic, 12 month sliding window, weighted)}
\label{fig:Results7}
\end{figure}
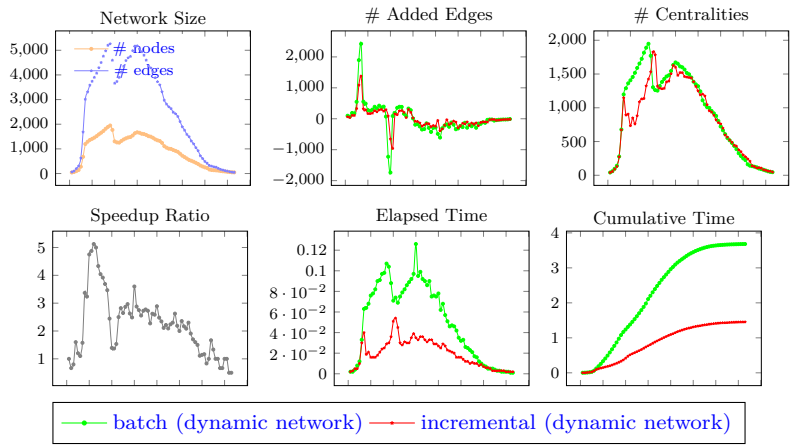

\begin{description}
    \item [\textbf{Unweighted Networks}:] Regarding the evaluation of full dynamic networks, the Autonomous systems AS-733 dataset with 733 snapshots was used. \figref{fig:Results2} -- Network Size, shows the evolution of the network concerning number of nodes and edges over the time. Please note that this setting contains removal of edges and therefore, \figref{fig:Results2} -- \# Added Edges, presents negative values (more removed edges that added for some snapshots). \figref{fig:Results2} -- \# Centralities, again shows that the batch version calculated centralities for all nodes in each snapshot,  while the incremental version did that only for affected nodes. It can be observed by this figure that the network has many edges and nodes variation causing significant change on the Elapsed Time required to compute centralities in each of the snapshots. Although this variation is also seen in the batch version (\figref{fig:Results2} -- Elapsed Time), the incremental version keeps elapsed times per snapshot lower, resulting in more efficient final execution times: 14,163 seconds for the incremental version, and the batch took 92,362 seconds. With this setup, we can see that the removal and addition of edges provokes an even more pronounced increase of speed-up ratio between the batch version and the dynamic version. The speed-up ratio achieves values of more than 15 times in some operations. In the second dynamic network setup, the evolution of the AS Caida Relationships Datasets over its 122 snapshots is shown in \figref{fig:Results4} -- Network Size. This network also has negative values for \# added edges, but it might be seen as a more stable network than the AS-733 dataset. Both number of affected centralities (\figref{fig:Results4} -- \# Centralities), and elapsed time per iteration (\figref{fig:Results4} -- Elapsed Time) are also presented. It can be observed that the removal and addition of edges, in this setup, provokes a pronounced increase of speed-up ratio between the batch version and the dynamic version. The speed-up ratio achieves values of up to 5 times in some operations with this network.\\

    \item [\textbf{Weighted Networks}:] Regarding the evaluation of full dynamic networks with the weighted version of the algorithm, two datasets were used. Starting this explanation with the DaysAll dataset,  \figref{fig:Results5} -- Network Size, shows the evolution of the network concerning number of nodes and edges over the time. Please note that this setting contains removal of edges and therefore, \figref{fig:Results5} -- \# Added Edges, presents negative values (more removed edges that added for some snapshots). \figref{fig:Results5} -- \# Centralities, again shows that the batch version calculated centralities for all nodes in each snapshot,  while the incremental version did that only for affected nodes. It can be observed by this figure that the network has many edges and nodes variation causing significant change on the Elapsed Time required to compute centralities in each of the snapshots. Although this variation is also seen in the batch version (\figref{fig:Results5} -- Elapsed Time), the incremental version keeps elapsed times per snapshot lower resulting in more efficient final execution times: 23,725 seconds for the incremental version while the batch took 67,357 seconds. The speed-up ratio achieves values of more than 4 times in some operations. In the second dynamic network setup, this time by using the evolution of the Bitcoinalpha Dataset over its 1919 daily snapshots is shown in \figref{fig:Results6} -- Network Size. This network also has negative values for \# added edges. Both number of affected centralities (\figref{fig:Results6} -- \# Centralities), and elapsed time per iteration (\figref{fig:Results6} -- Elapsed Time) are also presented. It can be observed that the removal and addition of edges, in this setup, provokes a pronounced increase of speed-up ratio between the batch version and the dynamic version. The speed-up ratio achieves values of up to 11 times in some operations with this network. In the third and last dynamic network setup, this time by using the evolution of the Bitcoinalpha Dataset with a 12 month snapshot setting, we have the results presented in \figref{fig:Results7}.  \figref{fig:Results7} -- Network Size again shows the increase and decrease of nodes and edges as the stream evolves and nodes and their connections are added or removed. Thus, this network also has negative values for \# added edges. Again, we present both number of affected centralities (\figref{fig:Results7} -- \# Centralities), and elapsed time per iteration (\figref{fig:Results7} -- Elapsed Time). Once again, it can be observed that the removal and addition of edges, in this setup, provokes an increase of speed-up ratio between the batch version and the dynamic version. The speed-up ratio achieves values above 5 times in some operations with this network.

\end{description}

\begin{figure}
  \centering
    
    \pgfplotstableread[col sep=semicolon]{step;number_of_nodes;number_of_edges
7;3042;5595
14;3074;5674
21;3100;5751
28;3130;5990
35;3172;6099
42;3181;6106
49;3203;6160
56;3216;6171
63;3237;6112
70;3260;6195
77;3292;6286
84;3315;6373
91;3333;6446
98;3359;6423
105;3389;6553
112;3442;6689
119;3415;6587
126;3452;6634
133;3470;6741
140;3505;6856
147;3537;6934
154;3576;6995
161;3595;7061
168;3620;7163
175;3627;7200
182;3641;7228
189;3670;7268
196;3693;7323
203;3720;7369
210;3717;7353
217;3752;7449
224;3761;7468
231;3797;7574
238;3812;7542
245;3836;7574
252;3864;7653
259;3877;7677
266;3932;7559
273;3966;7499
280;3991;7933
287;4023;7992
294;4061;8163
301;4088;8203
308;4134;8336
315;4144;8434
322;4185;8514
329;4185;8502
336;4206;8271
343;4227;8331
350;4264;8493
357;4301;8578
364;4332;8665
371;4351;8775
378;4370;8822
385;4404;8892
392;4416;8982
399;4452;9082
406;493;1234
413;545;1340
420;4502;9216
427;4532;9283
434;4582;9425
441;4599;9485
448;4653;9562
455;4675;9647
462;4719;9766
469;4744;9782
476;4772;9899
483;4799;9978
490;4843;10095
497;4885;10244
504;4904;10304
511;4936;10400
518;4969;10473
525;4999;10517
532;5029;10606
539;5080;10676
546;5132;10830
553;5176;10945
560;5198;10982
567;5219;11042
574;5277;11146
581;5333;11263
588;5361;11403
595;5402;11522
602;5437;11588
609;5471;11626
616;5514;11713
623;5568;11787
630;5598;11894
637;5615;11927
644;5666;12023
651;5705;12080
658;5738;12254
665;5816;12433
672;5855;12483
679;5900;12602
686;5939;12679
693;6127;13257
700;6209;13445
707;6301;13485
714;2086;4634
721;2080;4638
728;2122;4707
}\dataTableNetworksize
\pgfplotstableread[col sep=semicolon]{step;batch;dynamic
7;80;80
14;14;14
21;14;14
28;212;212
35;45;45
42;-15;-15
49;-7;-7
56;85;85
63;-27;-27
70;10;10
77;16;16
84;-35;-35
91;-15;-15
98;-97;-97
105;1;1
112;42;42
119;13;13
126;-14;-14
133;-27;-27
140;22;22
147;3;3
154;19;19
161;25;25
168;17;17
175;16;16
182;18;18
189;16;16
196;20;20
203;30;30
210;22;22
217;17;17
224;2;2
231;43;43
238;-35;-35
245;-75;-75
252;7;7
259;36;36
266;72;72
273;17;17
280;1;1
287;-38;-38
294;37;37
301;-4;-4
308;48;48
315;-12;-12
322;14;14
329;-35;-35
336;16;16
343;-10;-10
350;6;6
357;-11;-11
364;7;7
371;64;64
378;93;93
385;18;18
392;44;44
399;32;32
406;-7858;-7858
413;51;51
420;7872;7872
427;42;42
434;11;11
441;75;75
448;-20;-20
455;13;13
462;31;31
469;5;5
476;31;31
483;21;21
490;22;22
497;129;129
504;-36;-36
511;40;40
518;-11;-11
525;12;12
532;-71;-71
539;10;10
546;19;19
553;26;26
560;8;8
567;15;15
574;36;36
581;39;39
588;2;2
595;-6;-6
602;39;39
609;-73;-73
616;-6;-6
623;4405;4405
630;53;53
637;-2;-2
644;44;44
651;18;18
658;36;36
665;21;21
672;-15;-15
679;50;50
686;40;40
693;46;46
700;49;49
707;65;65
714;58;58
721;58;58
728;70;70
}\dataTableAddedEdges
\pgfplotstableread[col sep=semicolon]{step;batch;dynamic
7;3042;2008
14;3074;2272
21;3100;2125
28;3130;2445
35;3172;2342
42;3181;2069
49;3203;1451
56;3216;2119
63;3237;2323
70;3260;2400
77;3292;2306
84;3315;2371
91;3333;2463
98;3359;2397
105;3389;2249
112;3442;2434
119;3415;2410
126;3452;2352
133;3470;2493
140;3505;2261
147;3537;2414
154;3576;2461
161;3595;2458
168;3620;2552
175;3627;2127
182;3641;2309
189;3670;2430
196;3693;2583
203;3720;2739
210;3717;2550
217;3752;2643
224;3761;2479
231;3797;2665
238;3812;2182
245;3836;2631
252;3864;2541
259;3877;2431
266;3932;2505
273;3966;2311
280;3991;2675
287;4023;2488
294;4061;2593
301;4088;2624
308;4134;2872
315;4144;2762
322;4185;2893
329;4185;3036
336;4206;2737
343;4227;3010
350;4264;2500
357;4301;2694
364;4332;3025
371;4351;3268
378;4370;3185
385;4404;2875
392;4416;2921
399;4452;2735
406;493;4451
413;545;403
420;4502;4498
427;4532;2992
434;4582;3075
441;4599;2965
448;4653;3368
455;4675;3461
462;4719;3183
469;4744;3113
476;4772;3176
483;4799;3290
490;4843;3244
497;4885;3616
504;4904;3516
511;4936;3014
518;4969;2953
525;4999;3343
532;5029;3501
539;5080;3119
546;5132;2973
553;5176;3322
560;5198;3175
567;5219;3430
574;5277;3246
581;5333;3656
588;5361;3702
595;5402;3723
602;5437;4134
609;5471;3885
616;5514;4093
623;5568;5238
630;5598;3967
637;5615;4226
644;5666;3984
651;5705;3908
658;5738;4484
665;5816;4201
672;5855;4365
679;5900;4506
686;5939;3901
693;6127;4820
700;6209;3945
707;6301;4271
714;2086;1581
721;2080;1538
728;2122;1596
}\dataTableNumberCentralities
\pgfplotstableread[col sep=semicolon]{step;batch;dynamic
7;0.080;0.014
14;0.082;0.012
21;0.090;0.011
28;0.093;0.017
35;0.099;0.016
42;0.093;0.011
49;0.092;0.010
56;0.095;0.011
63;0.092;0.014
70;0.092;0.017
77;0.106;0.014
84;0.099;0.014
91;0.102;0.014
98;0.105;0.016
105;0.095;0.014
112;0.126;0.013
119;0.105;0.014
126;0.099;0.012
133;0.108;0.014
140;0.107;0.012
147;0.103;0.012
154;0.097;0.012
161;0.113;0.012
168;0.113;0.014
175;0.105;0.011
182;0.108;0.012
189;0.110;0.013
196;0.116;0.013
203;0.110;0.014
210;0.121;0.013
217;0.146;0.014
224;0.105;0.014
231;0.103;0.017
238;0.108;0.011
245;0.113;0.014
252;0.140;0.014
259;0.129;0.013
266;0.114;0.015
273;0.108;0.014
280;0.127;0.016
287;0.115;0.013
294;0.119;0.014
301;0.114;0.016
308;0.115;0.016
315;0.116;0.015
322;0.121;0.016
329;0.124;0.019
336;0.127;0.014
343;0.124;0.016
350;0.124;0.013
357;0.133;0.015
364;0.139;0.015
371;0.128;0.019
378;0.127;0.018
385;0.139;0.015
392;0.128;0.017
399;0.141;0.016
406;0.022;0.173
413;0.033;0.004
420;0.132;0.157
427;0.130;0.017
434;0.129;0.015
441;0.140;0.017
448;0.133;0.021
455;0.129;0.018
462;0.129;0.017
469;0.132;0.019
476;0.133;0.017
483;0.133;0.016
490;0.128;0.019
497;0.139;0.020
504;0.153;0.020
511;0.158;0.017
518;0.134;0.016
525;0.164;0.020
532;0.160;0.018
539;0.218;0.017
546;0.145;0.015
553;0.214;0.019
560;0.146;0.016
567;0.159;0.017
574;0.151;0.016
581;0.162;0.020
588;0.150;0.023
595;0.149;0.021
602;0.154;0.022
609;0.154;0.026
616;0.155;0.025
623;0.162;0.110
630;0.171;0.022
637;0.177;0.024
644;0.174;0.022
651;0.168;0.024
658;0.162;0.026
665;0.158;0.024
672;0.165;0.025
679;0.169;0.027
686;0.185;0.023
693;0.205;0.028
700;0.188;0.021
707;0.192;0.023
714;0.059;0.011
721;0.059;0.011
728;0.063;0.011
}\dataTableElapsedTime
\pgfplotstableread[col sep=semicolon]{step;batch;dynamic
7;0.588;0.171
14;1.152;0.258
21;1.761;0.35
28;2.445;0.441
35;3.082;0.532
42;3.79;0.617
49;4.474;0.699
56;5.13;0.783
63;5.808;0.874
70;6.453;0.966
77;7.178;1.053
84;7.852;1.147
91;8.531;1.242
98;9.24;1.339
105;9.982;1.449
112;10.726;1.537
119;11.428;1.635
126;12.173;1.736
133;12.948;1.827
140;13.742;1.911
147;14.496;2.001
154;15.28;2.113
161;16.011;2.204
168;16.876;2.305
175;17.653;2.398
182;18.406;2.491
189;19.192;2.584
196;20.025;2.736
203;20.889;2.827
210;21.751;2.921
217;22.611;3.016
224;23.367;3.107
231;24.114;3.222
238;24.848;3.322
245;25.666;3.421
252;26.536;3.522
259;27.358;3.617
266;28.138;3.719
273;28.957;3.824
280;29.845;3.93
287;30.787;4.035
294;31.672;4.142
301;32.527;4.253
308;33.393;4.361
315;34.282;4.462
322;35.15;4.575
329;36.038;4.692
336;36.926;4.799
343;37.774;4.901
350;38.623;5.004
357;39.564;5.109
364;40.485;5.217
371;41.465;5.331
378;42.41;5.439
385;43.299;5.546
392;44.236;5.651
399;45.151;5.764
406;45.948;6.023
413;46.126;6.053
420;46.407;6.232
427;47.368;6.354
434;48.283;6.463
441;49.236;6.575
448;50.16;6.7
455;51.17;6.826
462;52.112;6.955
469;53.088;7.077
476;54.14;7.197
483;55.122;7.317
490;56.098;7.44
497;57.143;7.569
504;58.313;7.7
511;59.416;7.828
518;60.419;7.96
525;61.449;8.092
532;62.5;8.235
539;63.567;8.389
546;64.561;8.514
553;65.734;8.677
560;66.788;8.809
567;67.881;8.95
574;68.949;9.089
581;70.01;9.23
588;71.076;9.373
595;72.235;9.699
602;73.359;9.845
609;74.549;10
616;75.681;10.166
623;76.742;10.489
630;77.874;10.642
637;79.052;10.794
644;80.026;11.602
651;81.169;11.753
658;82.389;11.913
665;83.591;12.084
672;84.759;12.246
679;85.935;12.41
686;87.209;12.574
693;88.515;12.783
700;89.551;13.387
707;90.869;13.585
714;91.21;13.927
721;91.635;14
728;92.082;14.062
}\dataTableCumulativeTime
\pgfplotstableread[col sep=semicolon]{step;speedup
7;5.71429
14;6.83333
21;8.18182
28;5.47059
35;6.1875
42;8.45455
49;9.2
56;8.63636
63;6.57143
70;5.41176
77;7.57143
84;7.07143
91;7.28571
98;6.5625
105;6.78571
112;9.69231
119;7.5
126;8.25
133;7.71429
140;8.91667
147;8.58333
154;8.08333
161;9.41667
168;8.07143
175;9.54545
182;9
189;8.46154
196;8.92308
203;7.85714
210;9.30769
217;10.4286
224;7.5
231;6.05882
238;9.81818
245;8.07143
252;10
259;9.92308
266;7.6
273;7.71429
280;7.9375
287;8.84615
294;8.5
301;7.125
308;7.1875
315;7.73333
322;7.5625
329;6.52632
336;9.07143
343;7.75
350;9.53846
357;8.86667
364;9.26667
371;6.73684
378;7.05556
385;9.26667
392;7.52941
399;8.8125
406;0.127168
413;8.25
420;0.840764
427;7.64706
434;8.6
441;8.23529
448;6.33333
455;7.16667
462;7.58824
469;6.94737
476;7.82353
483;8.3125
490;6.73684
497;6.95
504;7.65
511;9.29412
518;8.375
525;8.2
532;8.88889
539;12.8235
546;9.66667
553;11.2632
560;9.125
567;9.35294
574;9.4375
581;8.1
588;6.52174
595;7.09524
602;7
609;5.92308
616;6.2
623;1.47273
630;7.77273
637;7.375
644;7.90909
651;7
658;6.23077
665;6.58333
672;6.6
679;6.25926
686;8.04348
693;7.32143
700;8.95238
707;8.34783
714;5.36364
721;5.36364
728;5.72727
}\dataTableSpeedup

\pgfplotsset{footnotesize}
\begin{center}
%
%
\begin{tikzpicture}[scale=0.75]
\begin{axis}[
title={Network Size},
xticklabels={,,},
legend style={draw=none,legend to name=leg}
]
\addplot[smooth,mark=*,orange!50,mark size=0.75] table[y = number_of_nodes] from \dataTableNetworksize;
\addplot[smooth,mark=star,blue!50,mark size=0.75] table[y = number_of_edges] from \dataTableNetworksize;

   \addlegendentry{\# nodes}
   \addlegendentry{\# edges}

   \node [above left] (L) at (rel axis cs:0.7,0.6) {\ref{leg}};

\end{axis}
\end{tikzpicture}
\begin{tikzpicture}[scale=0.75]
\begin{axis}[
title={\# Added Edges},
xticklabels={,,},
]
\addplot[smooth,mark=*,green,mark size=0.75] table[y = batch] from \dataTableAddedEdges;
\addplot[smooth,mark=star,red,mark size=0.75] table[y = dynamic] from \dataTableAddedEdges;
\end{axis}
\end{tikzpicture}
\begin{tikzpicture}[scale=0.75]
\begin{axis}[
title={\# Centralities},
xticklabels={,,},
]
\addplot[smooth,mark=*,green,mark size=0.75] table[y = batch] from \dataTableNumberCentralities;
\addplot[smooth,mark=star,red,mark size=0.75] table[y = dynamic] from \dataTableNumberCentralities;
\end{axis}
\end{tikzpicture}
\\
\begin{tikzpicture}[scale=0.75]
\begin{axis}[
title={Speedup Ratio},
xticklabels={,,},
legend style={draw=none}
]
\addplot[smooth,mark=*,black!50,mark size=0.75] table[y = speedup] from \dataTableSpeedup;

\end{axis}
\end{tikzpicture}
\begin{tikzpicture}[scale=0.75]
\begin{axis}[
title={Elapsed Time},
xticklabels={,,},
]
\addplot[smooth,mark=*,green,mark size=0.75] table[y = batch] from \dataTableElapsedTime;
\addplot[smooth,mark=star,red,mark size=0.75] table[y = dynamic] from \dataTableElapsedTime;
\end{axis}
\end{tikzpicture}
\begin{tikzpicture}[scale=0.75]
\begin{axis}[
legend columns=-1,
legend entries={batch (dynamic network), incremental (dynamic network)},
legend to name=named,
title={Cumulative Time},
xticklabels={,,},
]
\addplot[smooth,mark=*,green,mark size=0.75] table[y = batch] from \dataTableCumulativeTime;
\addplot[smooth,mark=star,red,mark size=0.75] table[y = dynamic] from \dataTableCumulativeTime;
\end{axis}
\end{tikzpicture}
\\
\ref{named}
\end{center}

    \caption{Results for the as-733 
    (full dynamic, unweighted)}
\label{fig:Results2}
\end{figure}
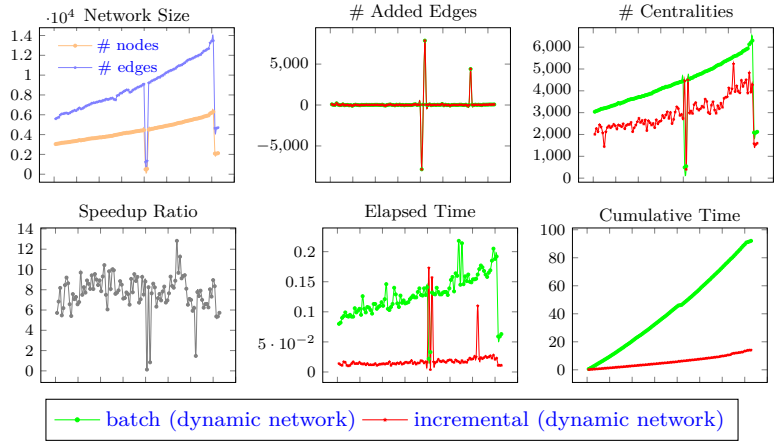

\begin{figure}[H]
  \centering

\pgfplotstableread[col sep=semicolon]{step;number_of_nodes;number_of_edges
1;16301;32955
2;16493;33372
3;16655;33340
4;16874;34335
5;17160;35013
6;17306;35547
7;17509;35829
8;17655;36070
9;17848;37172
10;18100;37497
11;18278;37559
12;18501;38265
13;18740;38501
14;18911;37991
15;19090;38562
16;19267;38930
17;19489;39859
18;19720;40198
19;19846;40485
20;20037;40474
21;20344;40941
22;20513;41069
23;20731;41762
24;20889;41820
25;21202;42925
26;21157;42734
27;21232;42872
28;21245;42481
29;21339;43283
30;21343;43425
31;21402;43616
32;21525;43969
33;21548;43892
34;21598;44245
35;21583;43780
36;21658;44308
37;21672;44327
38;21754;44322
39;21734;43902
40;21861;44829
41;21885;44472
42;21901;44588
43;22030;44839
44;22072;44780
45;22086;44576
46;22191;45086
47;22205;45001
48;22270;45410
49;22332;45392
50;22317;45196
51;22456;45050
52;22442;45550
53;22461;45148
54;22602;46085
55;22640;45738
56;22706;46413
57;22735;46167
58;22749;46085
59;22848;46578
60;22918;47090
61;22965;47193
62;23001;46864
63;23149;47368
64;23126;46779
65;23195;47534
66;23247;46606
67;23350;47242
68;23353;46655
69;23390;46095
70;23072;39342
71;23601;48006
72;23634;48202
73;23663;47777
74;23783;48648
75;23884;48746
76;23918;49089
77;24013;49332
78;24056;49391
79;24018;49196
80;24078;49056
81;20906;42994
82;24142;49200
83;24191;49225
84;24267;48986
85;24297;49583
86;24454;49830
87;24491;49826
88;24542;50183
89;24610;50738
90;24649;50462
91;24776;50827
92;24801;50881
93;24810;49958
94;24969;51238
95;24942;51145
96;25090;51178
97;25056;51106
98;25158;51234
99;25265;51755
100;25314;51444
101;25304;51510
102;25352;52048
103;25474;52317
104;25526;52412
105;25477;52363
106;25579;52546
107;25696;52666
108;25697;52762
109;25826;52401
110;25741;52423
111;25800;52383
112;25988;52365
113;26022;52691
114;8020;18203
115;26139;53156
116;26184;53289
117;26242;53174
118;26258;53601
119;26369;53231
120;26377;53255
121;26475;53381
}\dataTableNetworksize
\pgfplotstableread[col sep=semicolon]{step;batch;dynamic
1;32955;32955
2;417;417
3;-32;-32
4;995;995
5;678;678
6;534;534
7;282;282
8;241;241
9;1102;1102
10;325;325
11;62;62
12;706;706
13;236;236
14;-510;-510
15;571;571
16;368;368
17;929;929
18;339;339
19;287;287
20;-11;-11
21;467;467
22;128;128
23;693;693
24;58;58
25;1105;1105
26;-191;-191
27;138;138
28;-391;-391
29;802;802
30;142;142
31;191;191
32;353;353
33;-77;-77
34;353;353
35;-465;-465
36;528;528
37;19;19
38;-5;-5
39;-420;-420
40;927;927
41;-357;-357
42;116;116
43;251;251
44;-59;-59
45;-204;-204
46;510;510
47;-85;-85
48;409;409
49;-18;-18
50;-196;-196
51;-146;-146
52;500;500
53;-402;-402
54;937;937
55;-347;-347
56;675;675
57;-246;-246
58;-82;-82
59;493;493
60;512;512
61;103;103
62;-329;-329
63;504;504
64;-589;-589
65;755;755
66;-928;-928
67;636;636
68;-587;-587
69;-560;-560
70;-6753;-6753
71;8664;8664
72;196;196
73;-425;-425
74;871;871
75;98;98
76;343;343
77;243;243
78;59;59
79;-195;-195
80;-140;-140
81;-6062;-6062
82;6206;6206
83;25;25
84;-239;-239
85;597;597
86;247;247
87;-4;-4
88;357;357
89;555;555
90;-276;-276
91;365;365
92;54;54
93;-923;-923
94;1280;1280
95;-93;-93
96;33;33
97;-72;-72
98;128;128
99;521;521
100;-311;-311
101;66;66
102;538;538
103;269;269
104;95;95
105;-49;-49
106;183;183
107;120;120
108;96;96
109;-361;-361
110;22;22
111;-40;-40
112;-18;-18
113;326;326
114;-34488;-34488
115;34953;34953
116;133;133
117;-115;-115
118;427;427
119;-370;-370
120;24;24
121;126;126
;;-520
}\dataTableAddedEdges
\pgfplotstableread[col sep=semicolon]{step;batch;dynamic
1;16301;16301
2;16493;15803
3;16655;15967
4;16874;16442
5;17160;16572
6;17306;16801
7;17509;16929
8;17655;17056
9;17848;17441
10;18100;17447
11;18278;17571
12;18501;18045
13;18740;17930
14;18911;18352
15;19090;18318
16;19267;18586
17;19489;18638
18;19720;18961
19;19846;19188
20;20037;19293
21;20344;19786
22;20513;19751
23;20731;20014
24;20889;20044
25;21202;20132
26;21157;18962
27;21232;18964
28;21245;19367
29;21339;19585
30;21343;19203
31;21402;19146
32;21525;19454
33;21548;19512
34;21598;19678
35;21583;19797
36;21658;19804
37;21672;19716
38;21754;19825
39;21734;19990
40;21861;19905
41;21885;19900
42;21901;19917
43;22030;19913
44;22072;20091
45;22086;20128
46;22191;20280
47;22205;20270
48;22270;20424
49;22332;20497
50;22317;20664
51;22456;20595
52;22442;20775
53;22461;20806
54;22602;20843
55;22640;20798
56;22706;20907
57;22735;20811
58;22749;20921
59;22848;21005
60;22918;20988
61;22965;20686
62;23001;21017
63;23149;21253
64;23126;21110
65;23195;21121
66;23247;21137
67;23350;21213
68;23353;21674
69;23390;21779
70;23072;22623
71;23601;22688
72;23634;21926
73;23663;21802
74;23783;21797
75;23884;21683
76;23918;21550
77;24013;21057
78;24056;21263
79;24018;21604
80;24078;22011
81;20906;23307
82;24142;23404
83;24191;22120
84;24267;22193
85;24297;22363
86;24454;22062
87;24491;22372
88;24542;22382
89;24610;22435
90;24649;22395
91;24776;22342
92;24801;22305
93;24810;22572
94;24969;22954
95;24942;22809
96;25090;22828
97;25056;22886
98;25158;22731
99;25265;22959
100;25314;23048
101;25304;23037
102;25352;23453
103;25474;23383
104;25526;22995
105;25477;23213
106;25579;23435
107;25696;23374
108;25697;23444
109;25826;23728
110;25741;23463
111;25800;23458
112;25988;23488
113;26022;23451
114;8020;25997
115;26139;26100
116;26184;23622
117;26242;23664
118;26258;23726
119;26369;24099
120;26377;24272
121;26475;24110
;;24056
}\dataTableNumberCentralities
\pgfplotstableread[col sep=semicolon]{step;batch;dynamic
1;0.451;0.451
2;0.466;0.121
3;0.449;0.130
4;0.458;0.186
5;0.498;0.138
6;0.476;0.144
7;0.479;0.142
8;0.510;0.145
9;0.506;0.193
10;0.499;0.145
11;0.495;0.142
12;0.550;0.198
13;0.503;0.147
14;0.520;0.159
15;0.512;0.186
16;0.537;0.153
17;0.515;0.149
18;0.535;0.154
19;0.524;0.192
20;0.526;0.160
21;0.537;0.169
22;0.540;0.191
23;0.540;0.169
24;0.554;0.162
25;0.560;0.192
26;0.561;0.116
27;0.561;0.116
28;0.552;0.128
29;0.558;0.132
30;0.565;0.163
31;0.570;0.118
32;0.569;0.125
33;0.564;0.128
34;0.588;0.128
35;0.567;0.128
36;0.571;0.129
37;0.583;0.127
38;0.574;0.134
39;0.586;0.133
40;0.579;0.172
41;0.578;0.126
42;0.580;0.152
43;0.586;0.136
44;0.602;0.128
45;0.577;0.170
46;0.582;0.135
47;0.583;0.131
48;0.588;0.135
49;0.584;0.134
50;0.599;0.177
51;0.588;0.144
52;0.599;0.147
53;0.588;0.184
54;0.612;0.143
55;0.626;0.141
56;0.633;0.177
57;0.629;0.144
58;0.607;0.151
59;0.609;0.142
60;0.630;0.138
61;0.642;0.131
62;0.623;0.140
63;0.614;0.151
64;0.608;0.185
65;0.637;0.145
66;0.620;0.143
67;0.623;0.179
68;0.609;0.158
69;0.629;0.172
70;0.527;0.243
71;0.633;0.278
72;0.623;0.158
73;0.623;0.198
74;0.640;0.153
75;0.635;0.155
76;0.640;0.184
77;0.658;0.128
78;0.639;0.130
79;0.645;0.149
80;0.632;0.147
81;0.556;0.242
82;0.643;0.262
83;0.639;0.161
84;0.645;0.192
85;0.651;0.156
86;0.653;0.144
87;0.667;0.192
88;0.658;0.147
89;0.668;0.153
90;0.658;0.194
91;0.660;0.147
92;0.663;0.142
93;0.676;0.158
94;0.677;0.214
95;0.676;0.155
96;0.668;0.161
97;0.661;0.158
98;0.675;0.153
99;0.676;0.155
100;0.686;0.157
101;0.677;0.195
102;0.682;0.166
103;0.701;0.161
104;0.685;0.153
105;0.679;0.156
106;0.684;0.162
107;0.695;0.163
108;0.688;0.204
109;0.686;0.175
110;0.694;0.168
111;0.684;0.168
112;0.689;0.211
113;0.699;0.157
114;0.220;0.819
115;0.700;0.865
116;0.694;0.159
117;0.700;0.157
118;0.709;0.207
119;0.704;0.171
120;0.697;0.172
121;0.704;0.169
;;0.206
}\dataTableElapsedTime
\pgfplotstableread[col sep=semicolon]{step;batch;dynamic
1;0.451;0.451
2;0.917;0.572
3;1.366;0.702
4;1.824;0.888
5;2.322;1.026
6;2.798;1.17
7;3.277;1.312
8;3.787;1.457
9;4.293;1.65
10;4.792;1.795
11;5.287;1.937
12;5.837;2.135
13;6.34;2.282
14;6.86;2.441
15;7.372;2.627
16;7.909;2.78
17;8.424;2.929
18;8.959;3.083
19;9.483;3.275
20;10.009;3.435
21;10.546;3.604
22;11.086;3.795
23;11.626;3.964
24;12.18;4.126
25;12.74;4.318
26;13.301;4.434
27;13.862;4.55
28;14.414;4.678
29;14.972;4.81
30;15.537;4.973
31;16.107;5.091
32;16.676;5.216
33;17.24;5.344
34;17.828;5.472
35;18.395;5.6
36;18.966;5.729
37;19.549;5.856
38;20.123;5.99
39;20.709;6.123
40;21.288;6.295
41;21.866;6.421
42;22.446;6.573
43;23.032;6.709
44;23.634;6.837
45;24.211;7.007
46;24.793;7.142
47;25.376;7.273
48;25.964;7.408
49;26.548;7.542
50;27.147;7.719
51;27.735;7.863
52;28.334;8.01
53;28.922;8.194
54;29.534;8.337
55;30.16;8.478
56;30.793;8.655
57;31.422;8.799
58;32.029;8.95
59;32.638;9.092
60;33.268;9.23
61;33.91;9.361
62;34.533;9.501
63;35.147;9.652
64;35.755;9.837
65;36.392;9.982
66;37.012;10.125
67;37.635;10.304
68;38.244;10.462
69;38.873;10.634
70;39.4;10.877
71;40.033;11.155
72;40.656;11.313
73;41.279;11.511
74;41.919;11.664
75;42.554;11.819
76;43.194;12.003
77;43.852;12.131
78;44.491;12.261
79;45.136;12.41
80;45.768;12.557
81;46.324;12.799
82;46.967;13.061
83;47.606;13.222
84;48.251;13.414
85;48.902;13.57
86;49.555;13.714
87;50.222;13.906
88;50.88;14.053
89;51.548;14.206
90;52.206;14.4
91;52.866;14.547
92;53.529;14.689
93;54.205;14.847
94;54.882;15.061
95;55.558;15.216
96;56.226;15.377
97;56.887;15.535
98;57.562;15.688
99;58.238;15.843
100;58.924;16
101;59.601;16.195
102;60.283;16.361
103;60.984;16.522
104;61.669;16.675
105;62.348;16.831
106;63.032;16.993
107;63.727;17.156
108;64.415;17.36
109;65.101;17.535
110;65.795;17.703
111;66.479;17.871
112;67.168;18.082
113;67.867;18.239
114;68.087;19.058
115;68.787;19.923
116;69.481;20.082
117;70.181;20.239
118;70.89;20.446
119;71.594;20.617
120;72.291;20.789
121;72.995;20.958
;;21.164
}\dataTableCumulativeTime
\pgfplotstableread[col sep=semicolon]{step;speedup
1;1
2;3.85124
3;3.45385
4;2.46237
5;3.6087
6;3.30556
7;3.37324
8;3.51724
9;2.62176
10;3.44138
11;3.48592
12;2.77778
13;3.42177
14;3.27044
15;2.75269
16;3.5098
17;3.45638
18;3.47403
19;2.72917
20;3.2875
21;3.17751
22;2.82723
23;3.19527
24;3.41975
25;2.91667
26;4.83621
27;4.83621
28;4.3125
29;4.22727
30;3.46626
31;4.83051
32;4.552
33;4.40625
34;4.59375
35;4.42969
36;4.42636
37;4.59055
38;4.28358
39;4.40602
40;3.36628
41;4.5873
42;3.81579
43;4.30882
44;4.70312
45;3.39412
46;4.31111
47;4.45038
48;4.35556
49;4.35821
50;3.38418
51;4.08333
52;4.07483
53;3.19565
54;4.27972
55;4.43972
56;3.57627
57;4.36806
58;4.01987
59;4.28873
60;4.56522
61;4.90076
62;4.45
63;4.06623
64;3.28649
65;4.3931
66;4.33566
67;3.48045
68;3.85443
69;3.65698
70;2.16872
71;2.27698
72;3.94304
73;3.14646
74;4.18301
75;4.09677
76;3.47826
77;5.14062
78;4.91538
79;4.32886
80;4.29932
81;2.29752
82;2.4542
83;3.96894
84;3.35938
85;4.17308
86;4.53472
87;3.47396
88;4.47619
89;4.36601
90;3.39175
91;4.4898
92;4.66901
93;4.27848
94;3.16355
95;4.36129
96;4.14907
97;4.18354
98;4.41176
99;4.36129
100;4.36943
101;3.47179
102;4.10843
103;4.35404
104;4.47712
105;4.35256
106;4.22222
107;4.2638
108;3.37255
109;3.92
110;4.13095
111;4.07143
112;3.2654
113;4.45223
114;0.26862
115;0.809249
116;4.36478
117;4.4586
118;3.42512
119;4.11696
120;4.05233
121;4.16568
}\dataTableSpeedup

\pgfplotsset{footnotesize}
\begin{center}
%
%
\begin{tikzpicture}[scale=0.75]
\begin{axis}[
title={Network Size},
xticklabels={,,},
legend style={draw=none,legend to name=leg}
]
\addplot[smooth,mark=*,orange!50,mark size=0.75] table[y = number_of_nodes] from \dataTableNetworksize;
\addplot[smooth,mark=star,blue!50,mark size=0.75] table[y = number_of_edges] from \dataTableNetworksize;

   \addlegendentry{\# nodes}
   \addlegendentry{\# edges}

   \node [above left] (L) at (rel axis cs:0.7,0.6) {\ref{leg}};

\end{axis}
\end{tikzpicture}
\begin{tikzpicture}[scale=0.75]
\begin{axis}[
title={\# Added Edges},
xticklabels={,,},
]
\addplot[smooth,mark=*,green,mark size=0.75] table[y = batch] from \dataTableAddedEdges;
\addplot[smooth,mark=star,red,mark size=0.75] table[y = dynamic] from \dataTableAddedEdges;
\end{axis}
\end{tikzpicture}
\begin{tikzpicture}[scale=0.75]
\begin{axis}[
title={\# Centralities},
xticklabels={,,},
]
\addplot[smooth,mark=*,green,mark size=0.75] table[y = batch] from \dataTableNumberCentralities;
\addplot[smooth,mark=star,red,mark size=0.75] table[y = dynamic] from \dataTableNumberCentralities;
\end{axis}
\end{tikzpicture}
\\
\begin{tikzpicture}[scale=0.75]
\begin{axis}[
title={Speedup Ratio},
xticklabels={,,},
legend style={draw=none}
]
\addplot[smooth,mark=*,black!50,mark size=0.75] table[y = speedup] from \dataTableSpeedup;

\end{axis}
\end{tikzpicture}
\begin{tikzpicture}[scale=0.75]
\begin{axis}[
title={Elapsed Time},
xticklabels={,,},
]
\addplot[smooth,mark=*,green,mark size=0.75] table[y = batch] from \dataTableElapsedTime;
\addplot[smooth,mark=star,red,mark size=0.75] table[y = dynamic] from \dataTableElapsedTime;
\end{axis}
\end{tikzpicture}
\begin{tikzpicture}[scale=0.75]
\begin{axis}[
legend columns=-1,
legend entries={batch (dynamic network), incremental (dynamic network)},
legend to name=named,
title={Cumulative Time},
xticklabels={,,},
]
\addplot[smooth,mark=*,green,mark size=0.75] table[y = batch] from \dataTableCumulativeTime;
\addplot[smooth,mark=star,red,mark size=0.75] table[y = dynamic] from \dataTableCumulativeTime;
\end{axis}
\end{tikzpicture}
\\
\ref{named}
\end{center}

 \caption{Results for the as-Caida (full dynamic, unweighted)}
\label{fig:Results4}
\end{figure}
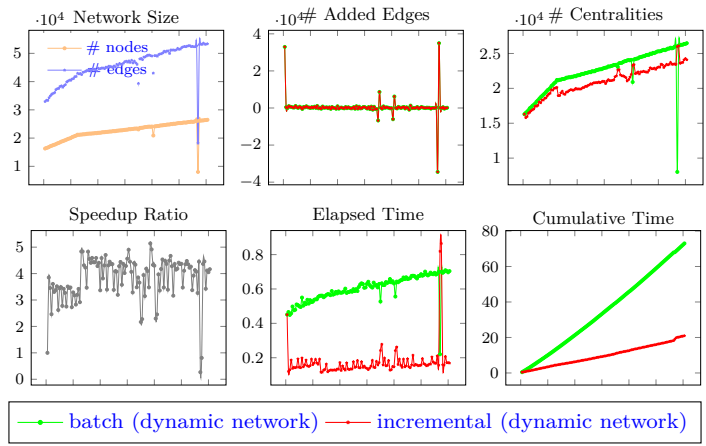

\subsection{Discussion}

\tableref{my-label-unweighted} shows that maximum values of speed-up for unweighted networks were obtained for step 122 in the High-energy physics (6,310 times), step 286 for AS 733 (17,909 times) and step 77 for AS-Caida (5,141 times). It can be seen that bigger speedups are obtained for small changes in the network where the number of recalculated centralities decreases significantly between snapshots. Regarding the initialization of the algorithm (step 1), depending on the initial network size, the incremental algorithm achieves the same speed-up for networks above 3k nodes / 5,5k edges, but for smaller networks, it can take more time than the batch algorithm in the first iteration. The degree of the affected nodes by addition or removal of nodes could also change the speed-up ratio, higher average degrees or dense networks reduce the speed-up ratio. Networks with high variability of addition and removal of nodes can also reduce the speed-up ratio. Finally, large or very large networks with smaller network changes between snapshots achieve the maximum speedup values. For unweighted networks, \tableref{my-label-unweighted} show that maximum values of speed-up were obtained in the Bitcoinalpha (11 times) in the step 1477, step 9 for DaysAll (4,5 times) and step 12 for Bitcoinalpha (5,125 times). It can be seen that bigger speedups are obtained for small changes in the network where the number of calculated centralities decreases significantly between snapshots. Regarding the initialization of the algorithm (step 1), depending on the initial network size, the incremental algorithm achieves reasonable speed-up, even for networks above 7k nodes / 48k edges. Again, the degree of the affected nodes by addition or removal of nodes could also change the speed-up ratio. We again conclude that higher average degrees or dense networks reduce the speed-up ratio, and networks with high variability of addition and removal of nodes can also reduce the speed-up ratio.

\begin{table}[]
\centering
\caption{Overview of the speed-up values achieved for the 7 datasets. Table shows the values for the initial step (full network for both algorithms), minimum, maximum and average. Apart from the speedup values, the conditions in terms of network size and number of added/removed edges per increment and the respective number of centralities calculated are also presented.}
\label{my-label-unweighted}
\begin{adjustbox}{max width=0.95\textwidth}
\begin{tabular}{llcccccccc}
\hline
\multicolumn{2}{c}{\multirow{2}{*}{}}                         & \multirow{2}{*}{\textbf{step}} & \multirow{2}{*}{\textbf{speedup}} & \multicolumn{2}{c}{\textbf{network size}} & \multicolumn{2}{c}{\textbf{centralities}} & \multicolumn{2}{c}{\textbf{added edges}} \\ \cline{5-10} 
\multicolumn{2}{c}{}                                          &                       &                          & \# nodes       & \# edges        & batch           & dynamic        & batch          & dynamic        \\ \hline
\multirow{4}{*}{\vtop{\hbox{\strut \textbf{High-energy physics}}\hbox{\strut {(Unweighted Incremental)}}}} & init:         & 1                     & 0,857                    & 4              & 2               & 4               & 4              & 2              & 2              \\ 
                                              & min:          & 2                     & 1,000                    & 9              & 6               & 9               & 7              & 4              & 4              \\ 
                                              & \textbf{max:} & \textbf{122}          & \textbf{6,310}           & \textbf{24023} & \textbf{280041} & \textbf{24023}  & \textbf{15023} & \textbf{3751}  & \textbf{3749}  \\ 
                                              & \multicolumn{2}{l}{average:}          & 3,910                    & 11784          & 110842          & 11784           & 7233           & 2594           & 2593           \\ \hline
\multirow{4}{*}{\vtop{\hbox{\strut \textbf{AS 733}}\hbox{\strut {(Unweighted Full Dynamic)}}}}              & init:         & 1                     & 0,989                    & 3015           & 5539            & 3015            & 3015           & 5539           & 5539           \\ 
                                              & min:          & 639                   & 0,027                    & 103            & 248             & 103             & 5629           & -11719         & -11719         \\ 
                                              & \textbf{max:} & \textbf{286}          & \textbf{17,909}          & \textbf{4020}  & \textbf{8030}   & \textbf{4020}   & \textbf{2268}  & \textbf{-1}    & \textbf{-1}    \\ 
                                              & \multicolumn{2}{l}{average:}          & 7,755                    & 4180           & 8533            & 4180            & 2919           & 11             & 11             \\ \hline
\multirow{4}{*}{\vtop{\hbox{\strut \textbf{AS Caida}}\hbox{\strut {(Unweighted Full Dynamic)}}}}            & init:         & 1                     & 1,000                    & 16301          & 32955           & 16301           & 16301          & 32955          & 32955          \\ 
                                              & min:          & 114                   & 0,269                    & 8020           & 18203           & 8020            & 25997          & -34488         & -34488         \\ 
                                              & \textbf{max:} & \textbf{77}           & \textbf{5,141}           & \textbf{24013} & \textbf{49332}  & \textbf{24013}  & \textbf{21057} & \textbf{243}   & \textbf{243}   \\ 
                                              & \multicolumn{2}{l}{average:}          & 3,818                    & 22518          & 45775           & 22518           & 20973          & 441            & 441            \\ \hline
\multirow{4}{*}{\vtop{\hbox{\strut \textbf{DaysAll}}\hbox{\strut {(Weighted Incremental)}}}}                & init:         & 1                     & 1,016                    & 2420           & 9318            & 2420             & 2420           & 9318           & 9318          \\ 
                                               & min:          & 2                    & 0,996                    & 4169           & 21396            & 4169             & 4115           & 12078          & 12078         \\ 
                                               & \textbf{max:} & \textbf{9}            & \textbf{4,174}           & \textbf{6741}  & \textbf{47283}  & \textbf{6741}    & \textbf{4114} & \textbf{353}   & \textbf{353}   \\ 
                                               & \multicolumn{2}{l}{average:}          & 3,239                    & 10044          & 94778           & 10044            & 8485           & 2242             & 2242             \\ \hline

\multirow{4}{*}{\vtop{\hbox{\strut \textbf{DaysAll}}\hbox{\strut {(Weighted Full Dynamic)}}}}               & init:         & 1                     & 1,047                    & 2420           & 9318            & 2420             & 2420           & 9318           & 9318          \\ 
                                               & min:          & 95                    & 0,892                    & 1153           & 2020            & 1153             & 1584           & -1124          & -1033         \\ 
                                               & \textbf{max:} & \textbf{9}            & \textbf{4,539}           & \textbf{6741}  & \textbf{47283}  & \textbf{6741}    & \textbf{4114} & \textbf{353}   & \textbf{353}   \\ 
                                               & \multicolumn{2}{l}{average:}          & 2,730                    & 7109           & 48295           & 7109            & 6443           & 21             & 21             \\ \hline
\multirow{4}{*}{\vtop{\hbox{\strut \textbf{Bitcoin-alpha-day}}\hbox{\strut {(Weighted Full Dynamic)}}}}    & init:         & 1                     & 1,000                        & 7              & 4               & 7               & 7              & 4              & 4              \\ 
                                               & min:          & 2                     & 0,333                    & 9              & 8               & 9               & 8              & 5              & 4              \\ 
                                               & \textbf{max:} & \textbf{1477}         & \textbf{11}              & \textbf{100}  & \textbf{118}     & \textbf{100}    & \textbf{8}  & \textbf{1}    & \textbf{1}         \\ 
                                               & \multicolumn{2}{l}{average:}          & 2,878                    & 181            & 250             & 181             & 91             & 1              & 1             \\ \hline
\multirow{4}{*}{\vtop{\hbox{\strut \textbf{Bitcoin-alpha-Month}}\hbox{\strut {(Weighted Full Dynamic)}}}} & init:         & 1                     & 1,000                    & 41             & 62              & 41              & 41             & 41             & 41          \\ 
                                               & min:          & 72                    & 0,500                    & 1518           & 4069            & 50              & 55             & -20            & -20         \\ 
                                               & \textbf{max:} & \textbf{12}           & \textbf{5,125}           & \textbf{1518}  & \textbf{4069}    & \textbf{1456}  & \textbf{849}    & \textbf{296}  & \textbf{296}   \\ 
                                               & \multicolumn{2}{l}{average:}          & 2,227                    & 934            & 2602            & 934             & 835            & 1              & 1            \\ \hline
\end{tabular}
\end{adjustbox}
\end{table}

\section{Conclusions and Future Work}\label{Conc}

In this paper, we presented an incremental and dynamic setup of the Laplacian Centrality algorithm. Through empiric experiments, we have shown that, the incremental and the dynamic setup of the algorithm are faster and more efficient than the batch version. Both settings, incremental and full dynamic, have shown improvements over the batch version of the algorithm, with both types of evolving networks (weighted and unweighted). Additionally, the dynamic algorithm can achieve a speed-up of more than 17 times when compared to the batch algorithm. The minimum speedup factor obtained in all tested scenarios was 4 times. This clearly shown the advantage of our Laplacian Centrality approach when dealing with evolving networks.

In the future, we plan to extend our research, by comparing the incremental setup of the algorithm with other incremental centrality measures like, for example, Betweenness Centrality and Closeness Centrality. For this purpose, we expect to improve results by achieving further optimization, both for weighted or unweighted networks.

\section*{Acknowledgements}
Part of this work is financed by the ERDF – European Regional Development Fund through the Operational Programme for Competitiveness and Internationalisation - COMPETE 2020 Programme  within project POCI-01-0145-FEDER-006961, and by National Funds through the FCT – Fundação para a Ciência e a Tecnologia (Portuguese Foundation for Science and Technology) as part of project  UID/EEA/50014/2013. Rui Portocarrero Sarmento also gratefully acknowledges funding from FCT (Portuguese Foundation for Science and Technology) through a PhD grant (SFRH/BD/119108/2016)

\end{document}